\documentclass[authoryear,12pt]{elsarticle}
\makeatletter
\def\ps@pprintTitle{%
 \let\@oddhead\@empty
 \let\@evenhead\@empty
 \def\@oddfoot{}%
 \let\@evenfoot\@oddfoot}
\makeatother
\usepackage{graphicx}
\usepackage[config=altsf]{subfig}
\usepackage{longtable}
\usepackage{blkarray}
\usepackage{amsmath,amssymb,amsfonts}
\usepackage{tensor}
\usepackage[margin=2.5cm]{geometry}
\usepackage[hypertexnames=false,final,colorlinks=false,pdfborder={0 0 0},bookmarksopen=true,bookmarksnumbered=true]{hyperref}
\usepackage{xspace}
\usepackage{booktabs}
\usepackage{siunitx}

\hypersetup{
  pdftitle={Mesoscopic behavior from microscopic Markov dynamics and its application to calcium release channels},
  pdfkeywords={calcium signaling \sep IP3 receptor \sep emergent properties of mesostates \sep Gillespie algorithm},
  pdfauthor={Nils Christian, Alexander Skupin, Silvia Morante, Karl Jansen, Giancarlo Rossi, Oliver Ebenhoeh},
}

\newcommand{\ie}{i.e.~}

\newcommand{\refeq}[1]{Eq.~\ref{#1}}
\newcommand{\refeqmain}[1]{Eq.~\ref{#1} in the main text}

\newcommand{\refsecmain}[1]{Sect.~\ref{#1} of the main text}

\newcommand{\refeqs}[1]{Eqs.~\ref{#1}}

\newcommand{\refeqn}[1]{\ref{#1}}%

\newcommand{\reffig}[1]{Fig.~\ref{#1}}

\newcommand{\reffign}[1]{\ref{#1}}
\newcommand{\Reffigure}[1]{Figure~\ref{#1}}%
\newcommand{\Reffigures}[1]{Figures~\ref{#1}}
\newcommand{\refsec}[1]{Sect.~\ref{#1}}

\newcommand{\reftable}[1]{Table~\ref{#1}}

\newcommand{\kij}[1]{k_{#1}}
\newcommand{\pei}[1]{p_{#1}}

\newcommand{\vij}[1]{v_{#1}}
\newcommand{\taui}[1]{\tau_{#1}}
\newcommand{\tausqi}[1]{\tau^{(2)}_{#1}}
\newcommand{\piij}[1]{\pi_{#1}}

\newcommand{\sooo}{s_{000}}
\newcommand{\sooi}{s_{001}}
\newcommand{\soio}{s_{010}}
\newcommand{\soii}{s_{011}}
\newcommand{\sioo}{s_{100}}
\newcommand{\sioi}{s_{101}}
\newcommand{\siio}{s_{110}}
\newcommand{\siii}{s_{111}}
\newcommand{\sijk}{s_{ijk}}
\newcommand{\sA}{\mathrm{A}}
\newcommand{\gA}{\mathcal{O}}
\newcommand{\globActRate}{a_{\mathcal{O}}}
\newcommand{\globDeactRate}{b_{\mathcal{O}}}

\newcommand{\lact}[1]{A_{#1}}
\newcommand{\ltomact}[2]{A_{[#1,#2]}}
\newcommand{\firstinbarVfromV}[2]{#1{:}#2}
\newcommand{\open}{O}
\newcommand{\closed}{C}
\newcommand{\topen}{\mathcal{T}(\open)}
\newcommand{\tclosed}{\mathcal{T}(\closed)}
\newcommand{\popen}{\mathcal{P}(\open)}

\newcommand{\pact}[1]{\mathcal{P}(\lact{#1})}
\newcommand{\tact}[1]{\mathcal{T}(\lact{#1})}
\newcommand{\isi}{\mathcal{T}(I)}
\newcommand{\isisq}{\mathcal{T}^{(2)}(I)}
\newcommand{\cvisi}{\mathrm{CV}\big(\isi\big)}

\newcommand{\Tgeneric}[2]{T\big(#1\big|#2\big)}
\newcommand{\TIgeneric}[3]{T\big(#1 {\rightarrow} #2\big|#3\big)}

\newcommand{\piijstarUV}[4]{P\big(#1 {\in} #3 {\rightarrow} #2 {\in} #4)}
\newcommand{\piijstarUUV}[4]{P\big(#1 {\in} #3 {\rightarrow} #2 {\in} #3 \wedge #3 {\rightarrow} #4)}
\newcommand{\piijstarUVZ}[5]{P\big(#1 {\in} #3 {\rightarrow} #2 {\in} #4 \wedge #3 {\rightarrow} #4 {\rightarrow} #5)}
\newcommand{\piijstarUUVZ}[5]{P\big(#1 {\in} #3 {\rightarrow} #2 {\in} #3 \wedge #3 {\rightarrow} #4 {\rightarrow} #5)}

\newcommand{\piiinUkinVcondUV}[4]{P\big(#1 {\in} #3 {\rightarrow} #2 {\in} #4 \big| #3 {\rightarrow} #4)}
\newcommand{\piiinUjinUcondUV}[4]{P\big(#1 {\in} #3 {\rightarrow} #2 {\in} #3 \big| #3 {\rightarrow} #4)}
\newcommand{\piiinUjinUcondUVZ}[5]{P\big(#1 {\in} #3 {\rightarrow} #2 {\in} #3 \big| #3 {\rightarrow} #4 {\rightarrow} #5)}

\newcommand{\piiinUjinUcondUkinV}[5]{P\big(#1 {\in} #3 {\rightarrow} #2 {\in} #3 \big| #3 {\rightarrow} #4 {\in} #5)}

\newcommand{\PiUVZ}[4]{P\big(#1 {\rightarrow} #2 {\rightarrow} #3 \big| #4 {\in} #1\big)}
\newcommand{\ZWUVYinU}[4]{T\big(#2 \big| #1 {\rightarrow} #2 {\rightarrow} #3 {\rightarrow} #4\big)}
\newcommand{\PWUVYiinU}[5]{P\big(#1 {\rightarrow} #5 {\in} #2\big|#1 {\rightarrow} #2 {\rightarrow} #3 {\rightarrow} #4\big)}
\newcommand{\XUVY}[4]{T\big(#1 \big| #4 {\in} #1 \wedge #1 {\rightarrow} #2 {\rightarrow} #3\big)}

\newcommand{\ThetaU}[2][U]{T\big(#1 \big| #2 \in #1\big)}
\newcommand{\ThetaUsq}[2]{T^{(2)}\big(#1 \big| #2 \in #1\big)}

\newcommand{\XUV}[3]{T\big(#1 \big| #3 \in #1 \wedge #1 {\rightarrow} #2\big)}
\newcommand{\XUVsq}[3]{T^{(2)}\big(#1 \big| #3 \in #1 \wedge #1 {\rightarrow} #2\big)}
\newcommand{\WUV}[4]{T\big(#1 \big| #3 \in #1 \wedge #1 {\rightarrow} #4 \in #2\big)}
\newcommand{\WUVsq}[4]{T^{(2)}\big(#1 \big| #3 \in #1 \wedge #1 {\rightarrow} #4 \in #2\big)}
\newcommand{\ZWUV}[3]{T\big(#2 \big| #1 {\rightarrow} #2 {\rightarrow} #3\big)}
\newcommand{\ZWUVsq}[3]{T^{(2)}\big(#2 \big| #1 {\rightarrow} #2 {\rightarrow} #3\big)}
\newcommand{\PU}[1]{P\big(#1\big)}
\newcommand{\PiUV}[3]{P\big(#1 {\rightarrow} #2 \big| #3 \in #1\big)}
\newcommand{\PiUVsubset}[4]{P\big(#1 {\rightarrow} #2 \big| #3 \in #4\big)}
\newcommand{\AWU}[3]{P\big(#1 {\rightarrow} #3 \in #2 \big| #1 {\rightarrow} #2\big)}
\newcommand{\CWUV}[4]{P\big(#1 {\rightarrow} #4 \in #2 \big| #1 {\rightarrow} #2 {\rightarrow} #3\big)}
\newcommand{\PWUV}[3]{P\big(#2 {\rightarrow} #3 \big| #1 {\rightarrow} #2\big)}
\newcommand{\PWUVbarVYiinU}[6]{P\big(#1 {\rightarrow} #6 \in #2\big|#1 {\rightarrow} #2 {\rightarrow} \firstinbarVfromV{#3}{#4} {\rightarrow} #5\big)}

\newcommand{\QUV}[4]{P\big(#1 {\rightarrow} #4 \in #2\big| #3 \in #1 \wedge #1 {\rightarrow} #2\big)}
\newcommand{\RUV}[4]{P\big(#1 {\rightarrow} #4 \in #2 \big| #3 \in #1\big)}
\newcommand{\PsiUVbarVY}[5]{P\big(#1 {\rightarrow} \firstinbarVfromV{#2}{#3} {\rightarrow} #4\big| #5 \in #1\big)}
\newcommand{\DWUVbarVY}[6]{P\big(#2 {\rightarrow} #6 \in #3\big| #1 {\rightarrow} #2 {\rightarrow} \firstinbarVfromV{#3}{#4} {\rightarrow} #5\big)}
\newcommand{\BWUV}[4]{P\big(#2 {\rightarrow} #4 \in #3\big| #1 {\rightarrow} #2 {\rightarrow} #3\big)}
\newcommand{\ZWUVbarVYinU}[5]{T\big(#2 \big| #1 {\rightarrow} #2 {\rightarrow} \firstinbarVfromV{#3}{#4} {\rightarrow} #5\big)}
\newcommand{\ZWUVbarVYinbarV}[5]{T\big(#4 \big| #1 {\rightarrow} #2 {\rightarrow} \firstinbarVfromV{#3}{#4} {\rightarrow} #5\big)}
\newcommand{\TIi}[3]{T\big(#1 {\rightarrow} #2\big|#3 \in #1\big)}
\newcommand{\TIisq}[3]{T^{(2)}\big(#1 {\rightarrow} #2\big|#3 \in #1\big)}

\newcommand{\ipt}{IP$_3$\xspace}
\newcommand{\iptm}{\textrm{IP}_3}
\newcommand{\iptr}{IP$_3$R\xspace}
\newcommand{\ca}{Ca$^{2+}$\xspace}
\newcommand{\cam}{\textrm{Ca}^{2+}}

\journal{Journal of Theoretical Biology}

\begin{document}

\begin{frontmatter}

\title{Mesoscopic behavior from microscopic Markov dynamics and its application to calcium release channels}

\author[1,2]{Nils Christian}
\ead{nils.christian@uni.lu}
\author[2,3]{Alexander Skupin}
\author[4]{Silvia Morante}
\author[5]{Karl Jansen}
\author[4]{Giancarlo Rossi}
\author[1,6]{Oliver Ebenh\"oh\corref{cor1}}
\ead{ebenhoeh@abdn.ac.uk}
\address[1]{University of Aberdeen, Department of Physics, Meston Walk, Aberdeen AB24 3UE, UK}
\address[2]{University Luxembourg, Luxembourg Centre for Systems Biomedicine, 7, avenue des Hauts-Fourneaux, L-4362 Esch-sur-Alzette, Luxembourg}
\address[3]{Institute for Systems Biology, 401 Terry Ave N, Seattle, WA, 98109, USA}
\address[4]{Universit\`a di Roma {\it Tor Vergata} and INFN, Sezione di Roma 2, Via della Ricerca Scientifica, I-00133 Roma, Italy}
\address[5]{NIC/DESY Zeuthen, Platanenallee 6, D-15738 Zeuthen, Germany}
\address[6]{Helmholtz Centre Potsdam, GFZ German Research Centre for Geosciences, 14473 Potsdam, Germany}
\cortext[cor1]{Corresponding author. Tel. +44 7580 078809}

\begin{abstract}
A major challenge in biology is to understand how molecular processes determine
phenotypic features. 
We address this fundamental problem in a class of model systems by
developing a general mathematical framework that allows the calculation of
mesoscopic properties from the knowledge of microscopic Markovian
transition probabilities.
We show how exact analytic formulae for the first and second moments
of resident time distributions in mesostates can be derived from
microscopic resident times and transition probabilities even for
systems with a large number of microstates. We apply our formalism
to models of the inositol trisphosphate receptor, which
plays a key role in generating
calcium signals triggering a wide variety of cellular responses.
We demonstrate how experimentally accessible quantities, such as opening and closing times
and the coefficient of variation of inter-spike intervals,
and other, more elaborated, quantities can be analytically calculated from the underlying
microscopic Markovian dynamics.
A virtue of our approach is that we do not need to follow the detailed
time evolution of the whole system, as we derive the relevant
properties of its steady state without having to take into account the often
extremely complicated transient features. We emphasize that our
formulae fully agree with results obtained by stochastic simulations
and approaches based on a full determination of the microscopic system's
time evolution.
We also illustrate how experiments can be devised to
discriminate between alternative molecular models of the inositol trisphosphate receptor.
The developed approach is applicable to any system described by a 
Markov process and, owing to the analytic nature of the
resulting formulae, provides an easy way to characterize also rare
events that are
of particular importance to understand the intermittency properties of complex dynamic systems.

\end{abstract}


\end{frontmatter}

\section{Introduction}
A major problem in biology is how characteristic features of living systems
can be explained as emergent properties
from underlying, elementary physico-chemical processes~\citep{Schrodinger1944}. 
The rapid technological advancements in the field of molecular biology allow
observation of phenomena
at many scales simultaneously. Thus, not surprisingly, 
in the last decades again a strong focus has been put on 
the occurrence of emerging properties in self-organizing systems,
the understanding of which is essential for a more complete picture of life~\citep{Laughlin2000}.

A fruitful approach to study complex systems, which include biological systems
as prominent examples, is by scale separation, separating the dynamics
on a microscopic scale from that on a
mesoscopic or macroscopic scale at which the elementary, microscopic
subunits are functionally organized~\citep{Mezard1987}.
The definition of what has to be considered as ``microscopic'' depends 
on the scale at which observations are made.
For molecular interaction 
networks, for example, elementary subunits can be atoms or molecules, 
the configuration of which define the mesoscopic state of a cell. In
population dynamics single cells or organisms should be considered as
elementary units, the interplay of which gives rise to the mesoscopic
dynamics on the population level.
This scale
separation allows decoupling the ``fast" dynamics of the
self-interacting elementary constituents from the ``slower" evolution
of the emergent mesoscopic degrees of freedom,
driven or influenced by external ``forces".

The notion of scale separation is very useful because it provides a
conceptually appealing and mathematically consistent way to
model the system at different complexity scales. At a
lower scale level the number of relevant degrees of freedom is
usually quite large and the network of their interactions is extremely
complicated.
Taking inspiration from the ``central limit theorem", one can hope to
be able to describe the system's underlying micro-dynamics by some
sort of simple stochastic process, most often of the Markov type, that
is intended to model the complicated network of individual molecular
interactions.
At
a higher and more structured level, new collective degrees of
freedom emerge from the more basic ones, that are in turn driven by 
external ``forces", such as temperature, pH, concentrations or gate
potentials, eventually back-reacting on the microscopic dynamics.
Of course many variations of this general scheme are
possible and have been explored in the literature with applications to
the description of channel activation~\citep{Shuai2007,Colquhoun1982}, neuronal
activity~\citep{Schwalger2010}, organelle interaction within the
cell~\citep{Heinrich2005}, homopolymer folding~\citep{LaPenna2004},
immunological response~\citep{Parisi1990} and many other processes.

In this paper we develop a general mathematical formalism able to
describe the micro-to-meso transition step, showing how from the
underlying microscopic
Markovian dynamics it is possible to derive fairly simple analytic
formulae characterizing and describing the emergence of higher level
mesoscopic properties. The key feature of the present approach is that
one can directly compute steady-state properties of mesostates, which we
define as any ensemble of microstates,
without the need of following the whole time-dependent transient
dynamics of the system. 
Our analytic formulae are directly derived from the underlying
Markovian dynamics by considering all possible paths of the system.
Despite the straight-forward nature of our approach, we are not aware
of any analogous method previously published.
An approach that is based on following in detail the dynamics of
micro- and mesostates (though the authors do not call them this way)
has been developed in the seminal paper of \citet{Colquhoun1982},
which is based on analysing the system dynamics in terms of the
Laplace-transform of the time evolution equations.
The time evolution of microstates is guided by the standard "master equation", 
that is in turn ruled by the elementary Markovian transition probabilities 
between pairs of microstates, while the time evolution of mesostates is driven 
by non-Markovian equations resulting from summing over the microstates 
defining each one of the mesostates.
Applications and the further development of this approach have been
reviewed in~\cite{Ball2000} and have been applied in a variant
in~\cite{Moenke2012}.
A notable difference of our method, compared to other 
approaches, is
that it provides an efficient way to compute the first statistical
moments of the observable mesoscopic process for an underlying
arbitrarily complex microscopic dynamics, as it avoids inversion of
the often extremely large transition probability matrices encoding the
elementary underlying Markov processes, by breaking them down into smaller sub-matrices. In this way one is in position
to more easily discriminate among different models by comparing
experimentally measurable moment relations with their analytic
expressions.

We illustrate our concepts and methods by studying the opening and closing dynamics of calcium driven channels
of which the so-called inositol trisphosphate receptor (\ipt receptor, or \iptr)~\citep{Allbritton92,Bezprozvanny94} is an
important and intensively studied biological prototype.
The \ipt pathway is a predominant release mechanism of calcium from intracellular 
stores and an important physiological second messenger pathway~\citep{Berridge98}.
The release is induced by moderate increase of the cytosolic calcium concentrations, 
known as calcium induced calcium release (CICR),
but inhibited by higher calcium concentrations~\citep{Bezprozvanny91}.
The release is countered by an ATP-dependent transport of free calcium back into 
storage compounds by SERCA pumps~\citep{MacLennan97}.
A spatially inhomogeneous distribution of calcium channels leads to a strong local coupling
with the consequence that the stochastic behavior of single channels influences the whole-cell 
behavior~\citep{skupin08,Skupin2010}. 
Thus, the \ipt receptors form the microscopic basis for the generation of complex mesoscopic calcium
signaling behavior~\citep{Berridge97,FalckeMalchow,Skupin2009}.

We select this particular example because a) it represents an important biological process
required for the generation of calcium signals, b) the underlying microscopic dynamics
can be described as a Markov process, c) mesostates and their statistical properties,
such as opening times, inter opening event intervals (inter-spike intervals) and their corresponding coefficients of variation (CV), are
important and experimentally measurable output quantities and d) more complex mesostates
can be defined whose properties are highly illustrative for
understanding how higher level properties emerge from the
underlying microscopic structure.
We first present in detail the theory of dwell times in
mesoscopic states, which allows analytic calculations of the first and second moments
of resident time distributions from the underlying Markovian dynamics of the microstates.
We then apply our approach to several alternative models of the \ipt receptor dynamics
and show how predicted moment relations can be used for model discrimination.

\section{Mesostates: definition and transition probabilities}
\label{sec:theory}
We consider a general Markovian system with $s$ \textit{microstates}. Let $\kij{mn}$ 
denote the rate constant for the elementary transition
from microstate $m$ to $n$ and $\pei{m}$ the stationary probability
of being in microstate $m$. Then the stationary probability rates, 
$v_{mn}$, are defined by
\begin{equation}
  \vij{mn}=\pei{m}\kij{mn}\, . \label{eq:VIJ}
\end{equation}

\begin{figure}[htb!]
  \centering
  \includegraphics[width=.35\linewidth]{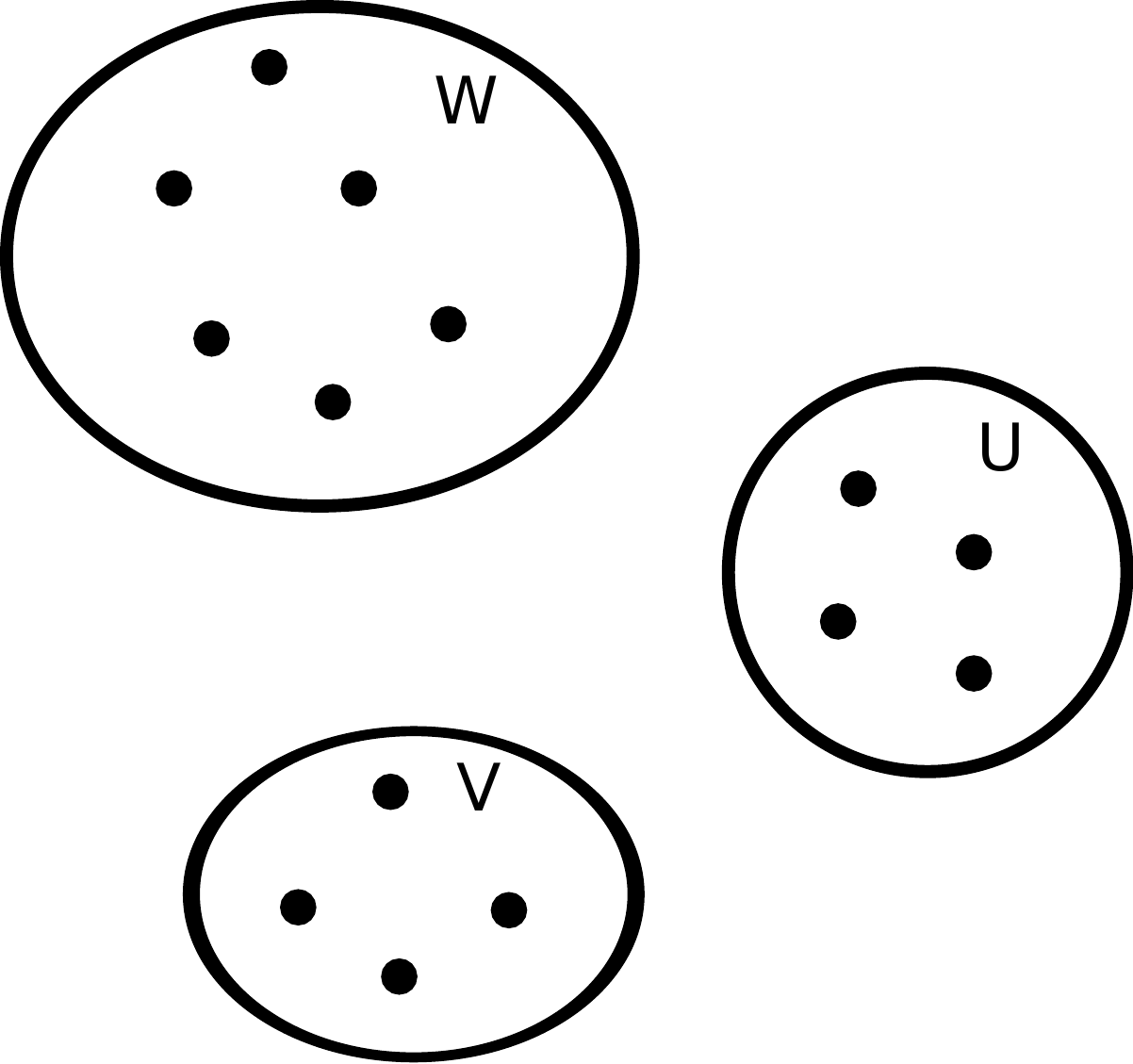}
  \caption{The configuration space of a Markov system. Small
    filled circles denote microstates. Sets of microstates enclosed by a curve 
    correspond to mesostates.}
  \label{fig:micronmeso}
\end{figure}
Stationarity means that the probability of being in a particular 
microstate remains constant in time. This requires that the 
stationary probability rates for transitions to a microstate must
be balanced by the rates for transitions from that state. 
This condition is expressed by the sum rule
\begin{equation}
  \sum\limits_{m} v_{mn} = \sum\limits_{l} v_{nl}. \label{eq:SUMR}
\end{equation}
Using \refeq{eq:VIJ}, the vector of stationary probabilities is obtained as 
a solution of the linear system
\begin{equation}
\sum\limits_{m} p_m (k_{mn} -  \delta_{mn}\sum\limits_l k_{ml}) = 0 \, .\label{eq:LINEQ}
\end{equation}
The corresponding mean dwell (also called ``residence'') time in a microstate is given by 
\begin{equation}
  \taui{m}=\frac{\pei{m}}{\sum\limits_{n'} \vij{mn'}}=\frac{1}{\sum\limits_{n'} \kij{mn'}} \ ,\label{eq:TAUS}
\end{equation}
in terms of which the transition probability between microstates $m$ and $n$
can be expressed as
\begin{equation}
  \label{eq:piij}
  \piij{mn}=
  \frac{\kij{mn}}{\sum\limits_{n'} \kij{mn'}}=\taui{m} \, \kij{mn} \, .
\end{equation}

We define \textit{mesostates} as sets of microstates (see \reffig{fig:micronmeso}).
If not stated otherwise, we assume that mesostates are finite and non-overlapping.
From the elementary (microscopic) quantities defined above we will derive
transition probabilities ($P$) between mesostates and dwell times ($T$) 
in mesostates, which are the essential properties of the process.
These quantities depend on the history of the system and it is therefore crucial to
take into account the possibility that certain transitions between 
micro and/or mesostates may have occurred before entering or are occurring after leaving
a particular mesostate.

We use the convention that mesostates are denoted by upper case
letters whereas microstates are denoted by lower case letters and
are indicated as subscripts, as already used in \refeqs{eq:VIJ}--\refeqn{eq:piij}.

\subsection{Probabilities for mesostate transitions}

A transition between two mesostates $U$ and $V$ is defined as any series of microstate transitions that
lead from one mesostate ($U$) to the other mesostate ($V$) without visiting a microstate not belonging 
to either $U$ or $V$.
To characterize the mesoscopic observable dynamics, we have to
calculate transition rates which are non-Markovian and therefore in
general depend on the history of the system. The history dependence of
the system is reflected by conditional probabilities that take the
different microscopic transition paths into account.

\begin{figure}[htb!]
  \centering
  \includegraphics[width=.35\linewidth]{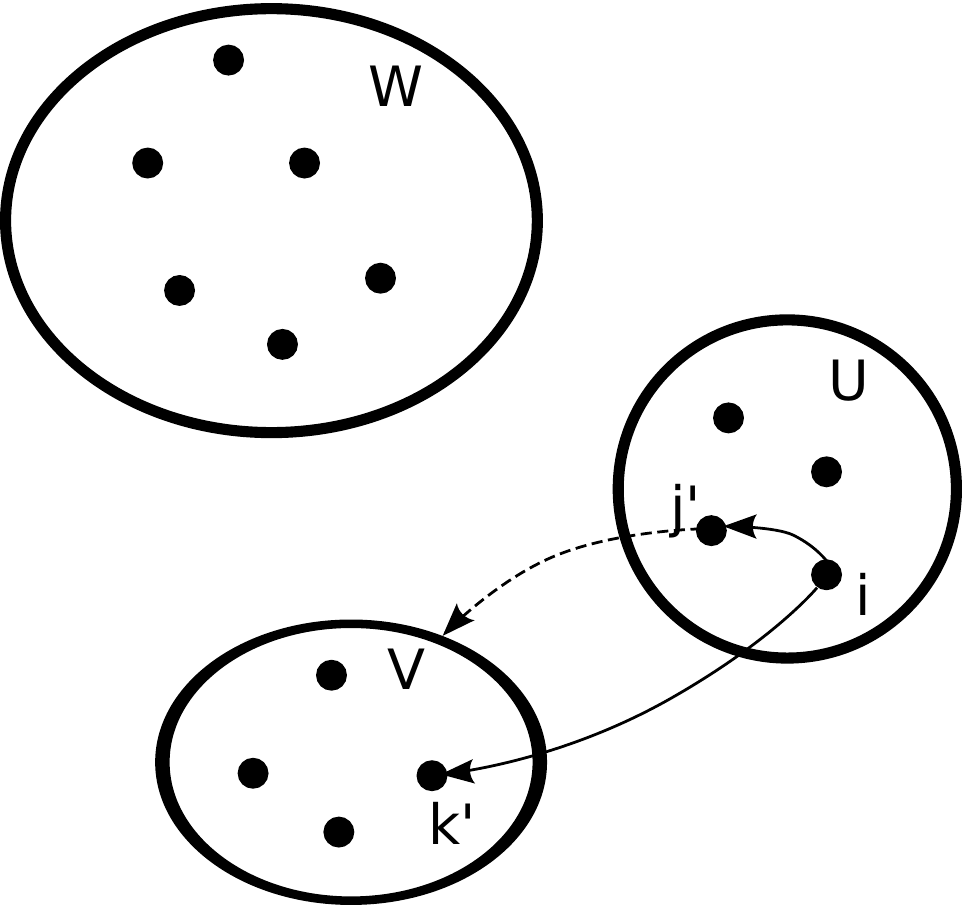}
  \caption{
    Different paths leading to a ${U{\rightarrow}V}$ mesostate
    transition starting from microstate ${i\in U}$.
    Two classes of possible events can occur: either the
    system visits another microstate $j'$ in $U$, or the system leaves
    directly to $k' \in V$}
  \label{fig:nextisvcondiinu}
\end{figure}

\paragraph{Probability that next mesostate is $V$ when system is in $i\in U$}
This quantity is the probability that for a system in a microstate $i$
of the mesostate $U$ the next mesostate transition will lead to $V$,
and is computed as 
the sum of the probability of leaving $U$
directly from $i$ to $V$ plus that of leaving to $V$ after freely moving
within $U$ (see \reffig{fig:nextisvcondiinu} for an illustration).
It can thus be calculated recursively by the formula
\begin{align}
  \PiUV{U}{V}{i}  &=\sum\limits_{k' \in V} \piij{ik'} + \sum\limits_{j' \in U} \piij{ij'} \PiUV{U}{V}{j'}   \label{eq:PUViinU}
    &=\sum\limits_{j' \in U} (\mathbf{1} - \pi)^{-1}_{ij'}\, \sum\limits_{k' \in V}\pi_{j'k'} \, , %
\end{align}
where $\pi=(\pi_{ij})$ is the matrix with coefficients $\pi_{ij}$.
The recursive nature of \refeq{eq:PUViinU} enables
for an efficient calculation of the first statistical moments of the
mesoscopic process generated by arbitrary complex microscopic
dynamics.

To introduce a notation that can be used for more complex transition
chains we write \refeq{eq:PUViinU} in the more
transparent form
\begin{align}
  \PiUV{U}{V}{i}  &=\sum_{k' \in V} \piijstarUV{i}{k'}{U}{V} \notag \\
  & \ + \sum_{j' \in U} \piijstarUUV{i}{j'}{U}{V} \, , \notag
\end{align}
where $\piijstarUUV{i}{j'}{U}{V}$ denotes the probability that the
system will undergo a direct transition from $i \in U$ to $j' \in U$
\emph{and} that the next visited mesostate is $V$ (this leads to the
second term of \refeq{eq:PUViinU}). Similarly,
$\piijstarUV{i}{k'}{U}{V}$ denotes the probability that the system
will undergo a direct transition from $i \in U$ to $k' \in V$, implying
that the next visited mesostate is $V$ (and therefore equals
$\piij{ik'}$).

\paragraph{Probability that first visited microstate in $V$ is $k$ when system is in $i\in U$ and the next mesostate is $V$}
With the help of \refeq{eq:PUViinU} we derive the probability that
the next mesostate transition from microstate $i \in U$ arrives in microstate $k \in V$,
under the condition that the next mesostate transition is $U
\rightarrow V$. Similar to \refeq{eq:PUViinU}, 
we get a recursive formula
\begin{align}
  \label{eq:quvikabstract}
  \QUV{U}{V}{i}{k}=& \piiinUkinVcondUV{i}{k}{U}{V} \notag \\
  & + \sum_{j \in U} \piiinUjinUcondUV{i}{j}{U}{V} \ \QUV{U}{V}{j}{k} \, ,
\end{align}
where the first summand describes a direct transition to $V$, without
visiting any other microstate within $U$, and can be calculated by
normalizing the unconditional probability $\piij{ik}$ with the
probability that the next mesostate is $V$
\begin{equation}
  \label{eq:piiinUkinVcondUV}
  \piiinUkinVcondUV{i}{k}{U}{V}=\frac{\piij{ik}}{\PiUV{U}{V}{i}} \, .
\end{equation}
The second summand takes into account the alternative process of
non-direct transitions in which the system visits another state $j\ne
i$ within $U$ before leaving to $V$.
With $M_{ij}$ defining the matrix elements (for a detailed derivation
of the second summand see Supplementary Material~{S1.1})%
\begin{equation}
  \label{eq:piiinUjinUcondUV}
  M_{ij}= \piiinUjinUcondUV{i}{j}{U}{V}= \left(1 - \frac{\sum\limits_{k' \in V}\piij{ik'}}{\PiUV{U}{V}{i}}\right)
  \frac{\piij{ij}\PiUV{U}{V}{j}}{\sum\limits_{i' \in U} \piij{ii'}\PiUV{U}{V}{i'}} \, ,
\end{equation}
the solution of the recursive formula \refeq{eq:quvikabstract} reads
\begin{equation}
  \label{eq:quvik}
  \QUV{U}{V}{i}{k} = \sum\limits_{j \in U}\left(\mathbf{1} - \mathbf{M}\right)^{-1}_{ij} \frac{\piij{jk}}{\PiUV{U}{V}{j}} \, .
\end{equation}

\paragraph{Probability that mesostate transition from $W$ to $U$ arrives in microstate $i\in U$}
To calculate the probability of arriving in $i \in U$, all stationary
probability rates from any state in $W$ to $i \in U$ are summed up and normalized
by all stationary probability rates from $W$ to any state $j'$ in $U$, leading to 
\begin{equation}
  \AWU{W}{U}{i}=\dfrac{\sum\limits_{g \in W} \pei{g}\kij{gi}}{\sum\limits_{g' \in W} \sum\limits_{j' \in U} \pei{g'}\kij{g'j'}} \, . \label{eq:P1}
\end{equation}
In contrast to expressions \refeqn{eq:PUViinU} and \refeqn{eq:quvik}, this
probability is conditional on a particular mesostate transition (\ie
not only on microstate properties). Therefore, it can only be
determined under the assumption of stationarity and depends on the
stationary probabilities.

\subsection{Transitions between mesostate subsets}
\label{sec:mesostatesubsets}
In some cases we might be interested in subsets of mesostates and
transition between them. To derive statistical properties of such a
subclass of mesostates, we need to know
probabilities and dwell times of
transition chains passing through mesostate $V$, where the first
visited microstate in $V$ is from a subset $V' {\subset} V$. We denote such a
transition chain by
\begin{equation}
  \dots \rightarrow \firstinbarVfromV{V'}{V} \rightarrow \dots \, .  \notag
\end{equation}
Determination of
the probability of the
transition chain $U \rightarrow \firstinbarVfromV{V'}{V} \rightarrow
Z$ when currently in $i \in U$ involves summing only over the
microstates of $V'$
\begin{equation}
  \label{eq:Pmesostatesubset}
  \PsiUVbarVY{U}{V'}{V}{Z}{i}= \sum_{k \in V'} \RUV{U}{V'}{i}{k} \ \PiUVsubset{V}{Z}{k}{V'} \, .
\end{equation}
The first term ensures that the first microstate in $V$ is from $V'$
(because the sum only includes microstates from $V'$), the second term
implicitly allows to move within the whole mesostate $V$ before the
system finally leaves to $Z$.

\section{Mesostate dwell times and raw moments}
\label{sec:MDTRM}

From the knowledge of the rate constants, $k_{mn}$, for microstate transitions,
we derive in this section analytic formulae for the average and
variance of dwell times in mesostates.
Because the dwell times are in general not Markovian variables, their
values will depend on the particular microstate transition by which the mesostate was first entered.
In general, the dwell times depend both on the microstate from which the previous mesostate was left
and on the microstate through which the current mesostate was entered.
We will denote dwell times in a mesostate by $\Tgeneric{U}{\dots}$.

The derivation of mesostate dwell times is not limited to a single
mesostate, but can also be performed for consecutive mesostates. For
example, the dwell time in the two consecutive mesostates $U$ and $V$
(\ie the dwell time in $U$ \emph{plus} the dwell time in the
subsequent $V$) will be denoted by $\TIgeneric{U}{V}{\dots}$.

\paragraph{Time spent in mesostate $U$ when system is in $i\in U$}
The time spent in mesostate $U$ from the moment the system first arrived in $i\in U$ 
is given by the recursive equation (see \refeq{eq:TAUS})
\begin{align}
  \ThetaU{i} &=\taui{i}+\sum\limits_{j' \in U} \piij{ij'} \ThetaU{j'} \notag
\end{align}
which leads to
\begin{align}  
  \label{eq:mesostatedwelltime}
  \ThetaU{i} &=\sum\limits_{j' \in U} (\mathbf{1} - \pi)^{-1}_{ij'}\, \taui{j'} \, .
\end{align}

To calculate the second raw moments of dwell times, one has to distinguish between 
consecutive events and alternative events which occur with certain
probabilities.
The microscopic dwell times of two consecutive events are independent random variables.
In general, the second raw moment of the sum of two independent random variables,
$X_a$ and $X_b$, is given by 
$  \langle (X_a + X_b)^2 \rangle
  = \langle X_a^2 \rangle + \langle X_b^2 \rangle + 2 \langle X_a \rangle \langle X_b \rangle .$
The second raw moment of the expected dwell time of two alternative processes is simply
the mean of the individual second raw moments weighted with their probabilities.
Thus, the second raw moment for the dwell time distribution in a mesostate~(\refeq{eq:mesostatedwelltime})
is
\begin{align}
  \ThetaUsq{U}{i} = & \left(1 - \sum\limits_{j' \in U} \piij{ij'}\right) \tausqi{i} \notag \\
  & \, + \sum\limits_{j' \in U} \piij{ij'}
  \left(\tausqi{i} + \ThetaUsq{U}{j'} + 2 \taui{i}\ \ThetaU{j'}\right) \notag \\
  = & \,\tausqi{i} + \sum\limits_{j' \in U} \piij{ij'} \left(\ThetaUsq{U}{j'} + 2 \taui{i}\ \ThetaU{j'}\right) \, ,
\end{align}
where $\tausqi{i}$ denotes the second raw moment of the microscopic dwell time $\taui{i}$.
The above formula simplifies if we assume that the microscopic dwell
times have an exponential probability distribution, so
that $\tausqi{i} = 2 \taui{i}^2$. In this case
\begin{align}
  \ThetaUsq{U}{i} = &\, 2 \taui{i}^2 + \sum\limits_{j' \in U} \piij{ij'} \left(\ThetaUsq{U}{j'} + 2 \taui{i}\ \ThetaU{j'}\right) \notag \\
  = & \,2 \taui{i} \underbrace{\Big(\taui{i} + \sum\limits_{j' \in U} \piij{ij'} \ThetaU{j'}\Big)}_{\ThetaU{i}} + \sum\limits_{j' \in U} \piij{ij'} \ThetaUsq{U}{j'} \, ,
\end{align}
leading to
\begin{align}
  \ThetaUsq{U}{i} = 2 \sum\limits_{j' \in U} (\mathbf{1} - \pi)^{-1}_{ij'}\, \taui{j'} \ThetaU{j'} \, .
  \label{eq:thetausq}
\end{align}

\paragraph{Time spent in mesostate $U$ when system is in $i \in U$ and the next visited mesostate is $V$}
The computation exactly parallels the line of arguments developed in
the derivation of \refeq{eq:quvik} and leads to the equation
\begin{align}
  & \XUV{U}{V}{i} = \taui{i}  + \sum\limits_{j \in U}
      \underbrace{\piiinUjinUcondUV{i}{j}{U}{V}}_{M_{ij}} \XUV{U}{V}{j}  \, ,\notag  
\end{align}
which has the solution
\begin{align}
  & \XUV{U}{V}{i} = \sum\limits_{j \in U}\left(\mathbf{1} - \mathbf{M}\right)^{-1}_{ij} \taui{j} \, .
  \label{eq:tUcondiinUandUtoV}
\end{align}
Similar to the derivation of \refeq{eq:thetausq} one finds for the second raw moment
\begin{align}
  & \XUVsq{U}{V}{i} = \frac{\sum\limits_{k' \in V}\piij{ik'}}{\PiUV{U}{V}{i}} \tausqi{i} \label{eq:RAW} \\
  & \quad + \sum\limits_{j \in U}
  \underbrace{\piiinUjinUcondUV{i}{j}{U}{V}}_{M_{ij}}
     \times\Big(\tausqi{i}+\XUVsq{U}{V}{j} + 2 \taui{i}\ \XUV{U}{V}{j}\Big)\, \notag .
\end{align}
For exponentially distributed $\taui{i}$, one gets
\begin{align}
  \XUVsq{U}{V}{i} =  2 \taui{i}^2 + \sum\limits_{j \in U} M_{ij} \Big(\XUVsq{U}{V}{j} + 2 \taui{i} \ \XUV{U}{V}{j} \Big)  \notag\\
  = 2 \taui{i} \underbrace{\big(\taui{i}+ \sum\limits_{j \in U} M_{ij} \XUV{U}{V}{j}\big)}_{\XUV{U}{V}{i}}
  + \sum\limits_{j \in U} M_{ij} \ \XUVsq{U}{V}{j}  \notag
\end{align}
finally leading to
\begin{align}  
  \XUVsq{U}{V}{i} &= 2 \sum\limits_{j \in U}\left(\mathbf{1} - \mathbf{M}\right)^{-1}_{ij} \taui{j} \XUV{U}{V}{j} \, .
\end{align}

\section{Application to models of the \iptr channel}
\label{sec:iptrmodels}
\iptr channels are the elementary building blocks
of calcium (\ca) signals~\citep{Berridge97} and thus play a crucial role as
physiological signal mediators.
Despite a lot of effort devoted to finding a mechanistic
interpretation of their properties~\citep{Young1992,Gin2009}, our
understanding is still rather poor.
We apply the general formalism developed in the previous sections to
three interesting \iptr models. 
We show how, from the basic processes defined by these models,
important mesoscopic physico-chemical properties of the system can be
obtained by our approach. These include
first and second raw moments of opening and closing time distributions and 
coherence (quantified by the coefficient of variation) of inter-spike intervals.
Comparison of theoretical results for such quantities with
experiments~\citep{Dellis2006,rahman2008} may allow to identify the
correct model of receptor structure and dynamics.
Similar channel properties have been subjected to theoretical
investigations~\citep{Higgins2009} aimed at a deeper understanding of
cell physiology.

\paragraph{Model descriptions}
\begin{sloppypar}
In the popular and comprehensive De~Young-Keizer model~\citep{Young1992} {\iptr}s are assumed to 
consist of four identical (independent) subunits, each endowed with three kinds of 
binding sites: a sensitizing \ipt binding site, an
activating and a deactivating \ca binding site. 
Since every site is either occupied or unoccupied, every subunit can reside in one
of 8 states. We therefore term this model the 8-state model.
The \iptr releases \ca
from the endoplasmic reticulum if at least three of the four subunits
are active, \ie if the \ipt and the activating \ca binding sites are
occupied and the deactivating \ca binding site is unoccupied. Due to
the release of \ca its concentration in the cytosol increases and can
cause the deactivation of the subunits by binding back to the deactivating site.
\end{sloppypar}

\begin{figure}[h]
  \centering
  \subfigure[8-state model]{
    \includegraphics[width=.4\textwidth]{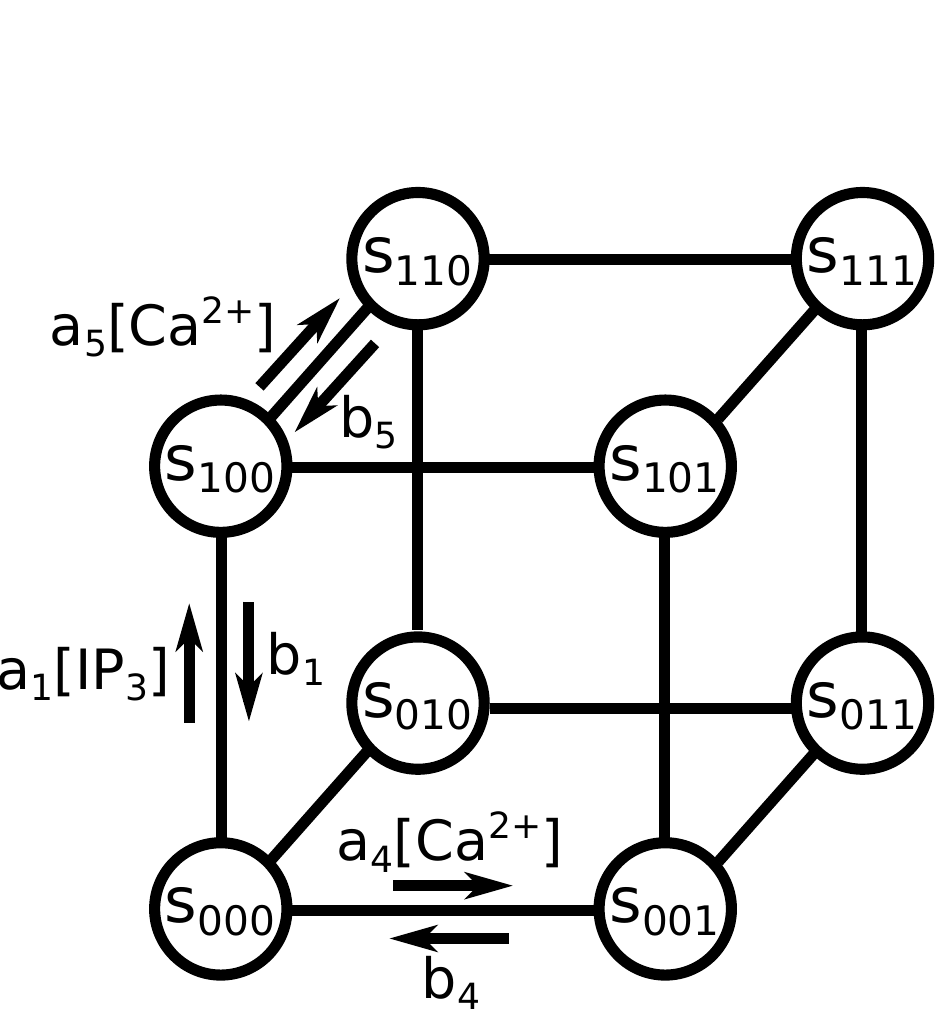}
  }
  \hspace{.05\textwidth}
  \subfigure[9-state model]{
    \includegraphics[width=.4\textwidth]{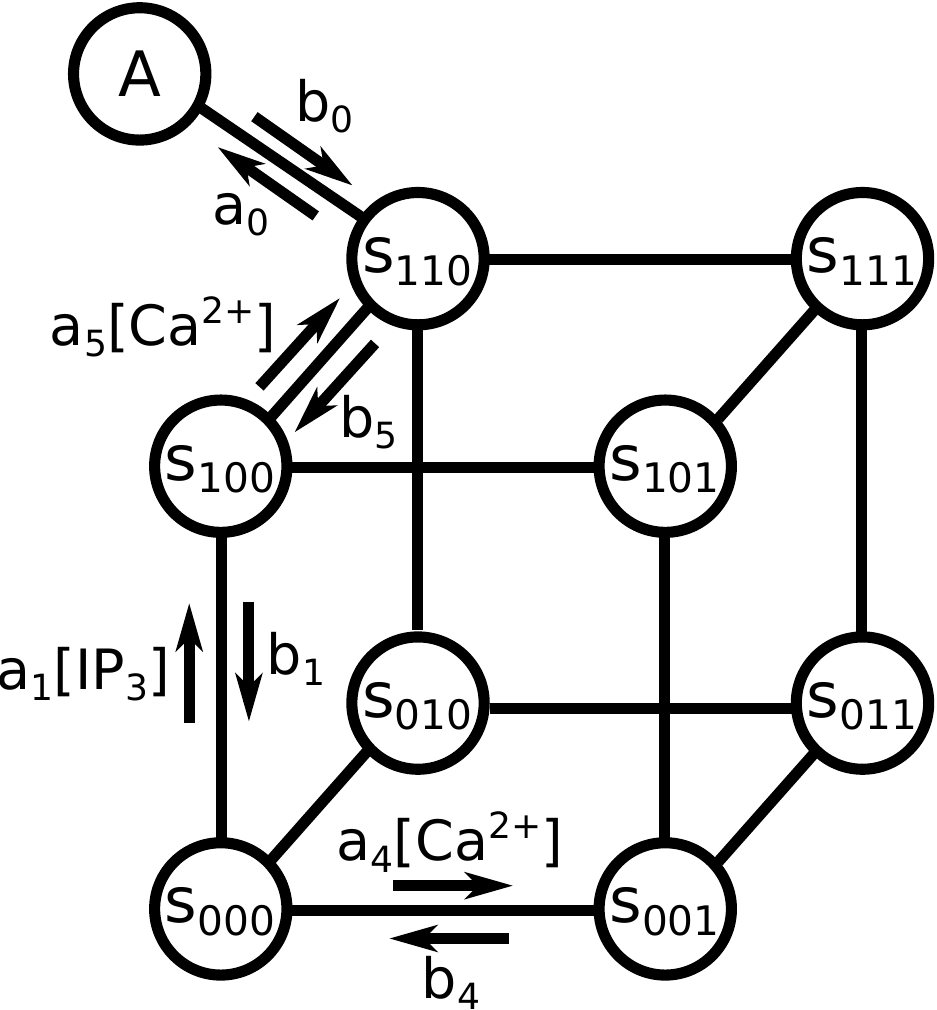}
  }
  \caption{Schematic representation of microstate transitions of a
    single \iptr subunits in the (a) 8-state and (b) 9-state model.
    Horizontal transitions 
    denote binding and release of
    \ca, vertical transitions denote binding and release of \ipt.
    The three binary digits denote the occupation state of the \ipt,
    activating and deactivating \ca binding site.
    In the 8-state model a subunit is active when it is in the state
    110. In the 9-state model a subunit is active when 
    from state 110 it undergoes a transition to the active state, A.}
  \label{fig:oldip3models}
\end{figure}

\citet{Shuai2007} have argued that with such a simple model some important
experimentally established results cannot be explained and
therefore proposed a model with a further activating state for each subunit, which we 
refer to as the 9-state model.
A key difference with the 8-state model is that in the 9-state model the
deactivation of a single subunit is independent on the ligand concentrations and allows for a fast modulation of the open state.
The main features of the two alternative models are
summarized in \reffig{fig:oldip3models}.

To calculate mesoscopic
properties of the models we are going to exploit the stochastic
analysis of probabilities and dwell times developed in the previous
sections.
To this end one has to first provide the rate constants
$k_{ij}$ between each pair of microstates characterizing the
model. The microstates are denoted by three binary digits
corresponding to the occupation of the \ipt, activating and
deactivating \ca binding site while A denotes the activated state. For a
single subunit in the 9-state model, the matrix $k$ reads
(see~\citet{Shuai2007})

{\footnotesize
  \begin{equation}
    \label{eq:MATR}
    k=
    \begin{blockarray}{rrrrrrrrrr}
      \sooo & \sooi & \soio & \soii & \sioo & \sioi & \siio & \siii & \sA\phantom{i} & \\
      \begin{block}{(rrrrrrrrr)r}
        0     & a_4 [\cam] & a_5 [\cam] & 0          & a_1 [\iptm] & 0           & 0           & 0           & 0\phantom{i}   & \sooo \\
        b_4 & 0            & 0          & a_5 [\cam] & 0           & a_3 [\iptm] & 0           & 0           & 0\phantom{i}   & \sooi \\
        b_5 & 0            & 0          & a_4 [\cam] & 0           & 0           & a_1 [\iptm] & 0           & 0\phantom{i}   & \soio \\
        0     & b_5        & b_4        & 0          & 0           & 0           & 0           & a_3 [\iptm] & 0\phantom{i}   & \soii \\
        b_1 & 0            & 0          & 0          & 0           & a_2 [\cam]  & a_5 [\cam]  & 0           & 0\phantom{i}   & \sioo \\
        0     & b_3        & 0          & 0          & b_2         & 0           & 0           & a_5 [\cam]  & 0\phantom{i}   & \sioi \\
        0     & 0          & b_1        & 0          & b_5         & 0           & 0           & a_2 [\cam]  & a_0\phantom{i} & \siio \\
        0     & 0          & 0          & b_3        & 0           & b_5         & b_2         & 0           & 0\phantom{i}   & \siii \\
        0     & 0          & 0          & 0          & 0           & 0           & b_0         & 0           & 0\phantom{i}   & \sA   \\
      \end{block}
    \end{blockarray}
    \ .
  \end{equation}
}

Naturally, in the absence of external fluxes, detailed-balance has to be obeyed by the elements of the 
matrix defined in \refeq{eq:MATR}, so that certain restrictions hold for the
parameters $a_i$ and $b_i$ (for details see~\citet{Shuai2007}).
A channel with four independent subunits can assume $9^4$ different microstates. 
Considering symmetries (\ie assuming that the subunits are indistinguishable), the transitions for the full four-subunit model can
be described by a $495\times 495$ matrix.
The single-subunit matrix for the 8-state model looks similar to the
9-state model, with the row/column related to the active state A
removed, as a subunit is considered already active if it is in state
${110}$. The $8^4$ different microstates of the four-subunit model can
be described by $330$ microstates when symmetries are taken into
account.

From a molecular perspective, the fast modulation of the channel opening may 
be more realistically considered as a whole molecule or protein phenomenon. Once 
the channel is in an ``excitable'' state with at least 3 active subunits, the 
protein may exhibit emergent dynamics leading to the final opening of 
the channel. Thus, we discuss as a third model a modification of the 8-state 
model which also displays a ligand-independent subunit deactivation. 
The difference with the models discussed above is, that instead of each subunit individually undergoing a
conformational change, we assume
that a channel opens by a joint conformational change of the four
subunits, locking the ligands until the channel
closes again.
Therefore we refer to this model as the global-activation model.

To describe this global conformation change, we have to introduce new
microstates reflecting open channels. 
In analogy to the 8-state model, 
we assume that conformation change is only possible if at least 3 subunits are
in state ${110}$. 
Channel opening and closing are described by the following additional microstate
transitions
\begin{align}
  4 \siio &\xrightarrow{\globActRate} \gA_4 &
  \gA_4  &\xrightarrow{\globDeactRate} 4 \siio \nonumber \\
  3 \siio + \sooo &\xrightarrow{\globActRate} \gA_{3}^{\sooo} &
  \gA_{3}^{\sooo} &\xrightarrow{\globDeactRate} 3 \siio + \sooo \nonumber \\
  3 \siio + \sooi &\xrightarrow{\globActRate} \gA_{3}^{\sooi} &
  \gA_{3}^{\sooi} &\xrightarrow{\globDeactRate} 3 \siio + \sooi \nonumber \\
  3 \siio + \soio &\xrightarrow{\globActRate} \gA_{3}^{\soio} &
  \gA_{3}^{\soio} &\xrightarrow{\globDeactRate} 3 \siio + \soio \nonumber \\
  3 \siio + \soii &\xrightarrow{\globActRate} \gA_{3}^{\soii} &
  \gA_{3}^{\soii} &\xrightarrow{\globDeactRate} 3 \siio + \soii \nonumber \\
  3 \siio + \sioo &\xrightarrow{\globActRate} \gA_{3}^{\sioo} &
  \gA_{3}^{\sioo} &\xrightarrow{\globDeactRate} 3 \siio + \sioo \nonumber \\
  3 \siio + \sioi &\xrightarrow{\globActRate} \gA_{3}^{\sioi} &
  \gA_{3}^{\sioi} &\xrightarrow{\globDeactRate} 3 \siio + \sioi \nonumber \\
  3 \siio + \siii &\xrightarrow{\globActRate} \gA_{3}^{\siii} &
  \gA_{3}^{\siii} &\xrightarrow{\globDeactRate} 3 \siio + \siii. \nonumber \\
\end{align}
Here, the symbol $\gA^{\sijk}_3$ denotes the open state with 3
subunits locked in the occupation state $\siio$ and the remaining
subunit locked in the occupation state $\sijk$. The symbol $\gA_4$
denotes the open state with all 4 subunits locked in the $\siio$
microstate. As in the 9-state model we assume that the rate
constants for activation ($\globActRate$) and deactivation
($\globDeactRate$) do not depend on the \ca and \ipt concentrations.
In total, this model has $8^4 + 8\cdot 4 + 1$ microstates which can be reduced
to 338 by taking symmetries into account.


\paragraph{Derivation of channel properties}
In the 8- and 9-state models, the channel is open if at least three
subunits are active. In the following we derive
channel opening and closing properties in the
general case where the number of channel subunits is $N$ and the
channel is open when at least $K_{th}\leq N$ of them are active. To
this end it is convenient to define the mesostates (with the integers
fulfilling $0<M<L\leq N$)

\nopagebreak[4]
\vspace{1em}
\nopagebreak[4]
\begin{tabular}{l p{.8\linewidth}}
  $\closed$ & set of microstates corresponding to a closed channel\\
  & \ (between $0$ and $K_{th}-1$ subunits are active)\\
  $\open$ & set of microstates corresponding to an open channel\\
  & \ ($K_{th}$ or more subunits are active)\\
  $\lact{L}$ & set of microstates with $L$ active subunits\\
  $\ltomact{M}{L}$ & set of microstates with $M$ to $L$ active subunits
\end{tabular}
\vspace{1em}

We further introduce the notation $\dots \rightarrow
\firstinbarVfromV{\lact{L}}{\ltomact{M}{L}} \rightarrow \dots$ to denote 
the fact that in a transition chain passing through $\ltomact{M}{L}$ 
the first visited microstate in that mesostate belongs to $\lact{L}$
(see \refsec{sec:mesostatesubsets}).

\paragraph{Opening and closing probabilities, mean opening time}

The basic probabilities for the channel being in an open or closed state are simply
\begin{equation}
  \PU{\open}=\sum\limits_{i \in \open} p_i \qquad \qquad
  \PU{\closed}=\sum\limits_{i \in \closed} p_i \ .
\label{eq:popenclosed}
\end{equation}
For the mean opening and closing time we have
\begin{equation}
\label{eq:meanopen}
\topen=\ZWUV{\closed}{\open}{\closed} \ ,
\end{equation}
\begin{equation}
\label{eq:meanclosed}
\tclosed=\ZWUV{\open}{\closed}{\open} \ .
\end{equation}

\paragraph{Time spent in an $\lact{L}$-related opening event}
An $\lact{L}$-related opening event is defined by the fact that a maximum
of $L$ active subunits with $L\in [K_{th},N]$ is reached during the
time the channel is open.
Although the dwell time and the corresponding probability are not directly measurable, their knowledge is
nevertheless interesting because understanding how
long a channel remains open in an $\lact{L}$-related opening event and how probable
such an event is may give useful insight on the extent the internal structure
of a channel influences its global characteristic opening behavior.
Such quantities are highly non-trivial %
but with our present approach, these and even more complicated quantities can be
derived systematically.
The probabilities and dwell times required here are similar or closely
follow the derivation described in the general section introducing
the mesostate subsets.

\begin{figure}[htb!]
  \centering
  \includegraphics[width=.4\linewidth]{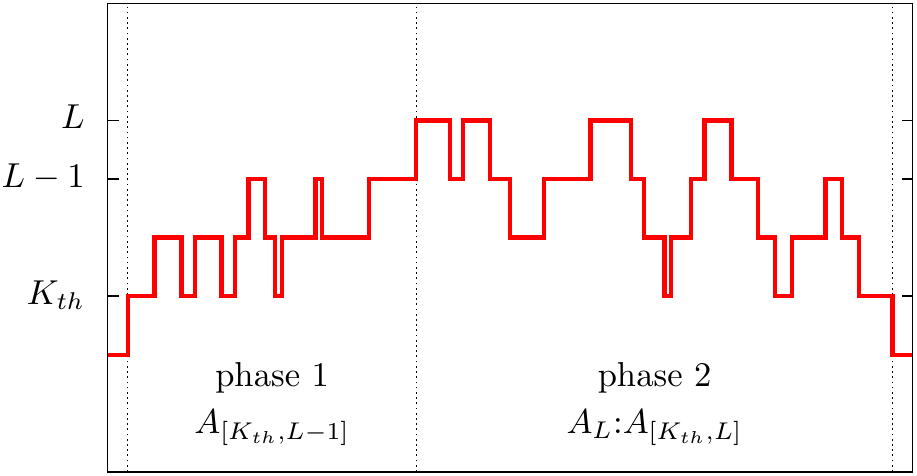}
  \caption{The time spent in an $\lact{L}$-related opening event can be
    divided into two parts. During phase~1 a maximum of $L-1$
    subunits are active. Phase~2 begins when for the first
    time $L$ subunits are active and ends when the channel closes.}
  \label{fig:lrelated}
\end{figure}

First the dwell time in the mesostate
$\ltomact{K_{th}}{L-1}$ has to be calculated under the condition that from this mesostate only
states with a maximal number of $L$ active subunits are visited before
the system reaches the closed state $\closed$ (phase~1 in
\reffig{fig:lrelated}). For this we need the probability of the transition chain
$\ltomact{K_{th}}{L-1} \rightarrow \firstinbarVfromV{\lact{L}}{\ltomact{K_{th}}{L}}
\rightarrow \closed$ when currently in $i \in \ltomact{K_{th}}{L-1}$, which is
(see \refeq{eq:Pmesostatesubset})
\begin{align}
  & \PsiUVbarVY{\ltomact{K_{th}}{L-1}}{\lact{L}}{\ltomact{K_{th}}{L}}{\closed}{i}= \notag \\
  & \quad \sum\limits_{l \in \lact{L}} \RUV{\ltomact{K_{th}}{L-1}}{\lact{L}}{i}{l} \ \PiUV{\ltomact{K_{th}}{L}}{\closed}{l} \, .
\end{align}
Using this expression, the probability that the first microstate in
$\ltomact{K_{th}}{L-1}$ is $i$, given the transition chain ${\closed \rightarrow
\ltomact{K_{th}}{L-1} \rightarrow \firstinbarVfromV{\lact{L}}{\ltomact{K_{th}}{L}} 
\rightarrow \closed}$, can be written in the form (for details see Eq.~{S1.18} in the Supplementary Material S1)%
\begin{align}
  \label{eq:probPhase1}
  & \PWUVbarVYiinU{\closed}{\ltomact{K_{th}}{L-1}}{\lact{L}}{\ltomact{K_{th}}{L}}{\closed}{i}=
  \frac{P_i^{\textrm{phase 1}}}{\sum\limits_{i' \in \ltomact{K_{th}}{L-1}} P_{i'}^{\textrm{phase 1}}} \, ,
\end{align}
where
\begin{align}
  P_i^{\textrm{phase 1}}=
  & \AWU{\closed}{\ltomact{K_{th}}{L-1}}{i} \notag \\
  & \quad \times \PsiUVbarVY{\ltomact{K_{th}}{L-1}}{\lact{L}}{\ltomact{K_{th}}{L}}{\closed}{i} \, . \notag
\end{align}
The time spent in phase~1 is therefore (see Eq.~{S1.17} in the Supplementary Material S1)%
\begin{align}
  \label{eq:talPhase1}
  & \ZWUVbarVYinU{\closed}{\ltomact{K_{th}}{L-1}}{\lact{L}}{\ltomact{K_{th}}{L}}{\closed} = \notag \\
  & \qquad \sum\limits_{i \in \ltomact{K_{th}}{L-1}}
   \PWUVbarVYiinU{\closed}{\ltomact{K_{th}}{L-1}}{\lact{L}}{\ltomact{K_{th}}{L}}{\closed}{i} \notag \\
        & \qquad \ \phantom{ \sum\limits_{i \in \ltomact{K_{th}}{L-1}}} \ \times \XUVY{\ltomact{K_{th}}{L-1}}{\firstinbarVfromV{\lact{L}}{\ltomact{K_{th}}{L}}}{\closed}{i}\, .
\end{align}
For computing the dwell time in phase~2 the conditional probability to arrive in 
microstate $l \in \lact{L} \subset \ltomact{K_{th}}{L}$, knowing that the next 
mesostate is closed (i.e.\ $\closed$), is needed. In analogy with \refeq{eq:probPhase1} this probability can be derived to be 
\begin{align}
  \label{eq:probPhase2}
  & \DWUVbarVY{\closed}{\ltomact{K_{th}}{L-1}}{\lact{L}}{\ltomact{K_{th}}{L}}{\closed}{l}=
  \frac{P_l^{\textrm{phase 2}}}{\sum\limits_{l' \in \lact{L}} P_{l'}^{\textrm{phase 2}}} \, ,
\end{align}
with
\begin{align}
  P_l^{\textrm{phase 2}}=
  & \ \BWUV{\closed}{\ltomact{K_{th}}{L-1}}{\lact{L}}{l} \PiUV{\ltomact{K_{th}}{L}}{\closed}{l} \, . \notag
\end{align}
The time spent in phase~2 is therefore
\begin{align}
\label{eq:talPhase2}
  & \ZWUVbarVYinbarV{\closed}{\ltomact{K_{th}}{L-1}}{\lact{L}}{\ltomact{K_{th}}{L}}{\closed} = \notag \\
  & \qquad \sum\limits_{l \in \lact{L}} \DWUVbarVY{\closed}{\ltomact{K_{th}}{L-1}}{\lact{L}}{\ltomact{K_{th}}{L}}{\closed}{l} \\
  & \qquad \phantom{\sum\limits_{l \in \lact{L}}} \ \times \XUV{\ltomact{K_{th}}{L}}{\closed}{l} \ . \notag
\end{align}
Altogether, summing \refeqs{eq:talPhase1} and~\refeqn{eq:talPhase2}, the mean time spent in an $\lact{L}$-related opening event is
\begin{align}
  \label{eq:tal}
  \tact{L}= &\ZWUVbarVYinU{\closed}{\ltomact{K_{th}}{L-1}}{\lact{L}}{\ltomact{K_{th}}{L}}{\closed} \notag \\
  &+ \ZWUVbarVYinbarV{\closed}{\ltomact{K_{th}}{L-1}}{\lact{L}}{\ltomact{K_{th}}{L}}{\closed}\, .
\end{align}

\paragraph{Probability of an $\lact{L}$-related opening event}

We now seek for an expression for the probability of an
$\lact{L}$-related opening event, $\pact{L}$, \ie for the fraction of
the opening events with a maximum of $L$ active subunits. To this end,
we first have to compute the auxiliary probability of leaving
$\ltomact{K_{th}}{L}$ to the closed state (\ie without visiting
$\lact{L+1}$), under the condition that we reached
$\ltomact{K_{th}}{L}$ from the closed state.
For this auxiliary probability one finds
\begin{equation}
  \sum\limits_{l' \in \lact{L}} \BWUV{\closed}{\ltomact{K_{th}}{L-1}}{\lact{L}}{l'} \PiUV{\ltomact{K_{th}}{L}}{\closed}{l'} \, .
\end{equation}
To obtain $\pact{L}$ this has to be multiplied by the probability of arriving 
in $L$, finally leading to
\begin{align}
  \label{eq:pal}
  \pact{L}=&\PWUV{\closed}{\ltomact{K_{th}}{L-1}}{\lact{L}} \\
  & \ \times \sum\limits_{l' \in \lact{L}} \BWUV{\closed}{\ltomact{K_{th}}{L-1}}{\lact{L}}{l'} \PiUV{\ltomact{K_{th}}{L}}{\closed}{l'} \notag .
\end{align}

\paragraph{Inter-spike interval}
An important measurable quantity is the so-called inter-spike interval, $\isi$, 
i.e.\ the interval between two opening events. It can be calculated by the 
weighted average over all the dwell times in $\open$ leading to a specific microstate in
$\closed$, to which one has to add the subsequent dwell time in $\closed$. This leads to the expression
\begin{align}
\label{eq:ti}
  & \TIi{\open}{\closed}{i}= \notag \\
  & \ \sum\limits_{j \in \closed} \Big(\QUV{\open}{\closed}{i}{j}  \big[\WUV{\open}{\closed}{i}{j} + \XUV{\closed}{\open}{j}\big]\Big) \ .
\end{align}
The associated second raw moment is consequently
\begin{align}
\label{eq:titwo}
  & \TIisq{\open}{\closed}{i}= \notag \\
  & \ \sum\limits_{j \in \closed} \Big(\QUV{\open}{\closed}{i}{j} \big[\WUVsq{\open}{\closed}{i}{j} + \XUVsq{\closed}{\open}{j} \\
  & \phantom{\ \sum\limits_{j \in \closed}(\QUV{\open}{\closed}{i}{j}[} \ + 2\WUV{\open}{\closed}{i}{j} \ \XUV{\closed}{\open}{j}\big] \Big) \ . \notag
\end{align}
To get quantities that are independent of the microstate $i$, \refeqs{eq:ti} and~\refeqn{eq:titwo}
have to be weighted with the probability of arriving in the
microstate $i \in \open$. One gets in this way for the inter-spike interval
\begin{equation}
  \isi=\sum\limits_{i \in \open}\AWU{\closed}{\open}{i}\,\TIi{\open}{\closed}{i}\, ,
\end{equation}
and for its second raw moment
\begin{equation}
  \isisq=\sum\limits_{i \in \open}\AWU{\closed}{\open}{i} \ \TIisq{\open}{\closed}{i} \, .
\end{equation}
From \refeqs{eq:ti} and~\refeqn{eq:titwo} one can construct the coefficient of variation
\begin{equation}
  \label{eq:isicv}
  \cvisi=\frac{\big(\isisq-(\isi)^2\big)^{1/2}}{\isi} \ .
\end{equation}

\section{Mesoscopic properties of the \iptr models}
\label{sec:iptrresults}

In this section, using the analytical formulae derived in
previous sections, we show results for
the behavior of some of the observables we have defined before as
function of the \ca and/or \ipt concentrations for the three models of
{\iptr}s. %
All our analytic results are checked against Gillespie-type simulations~\citep{Gillespie1977}.

In all the plots solid curves correspond to the results obtained with the
analytical approach, while isolated symbols
denote Gillespie simulation points. Error bars on the latter denote
standard error of the mean. In most cases, 
the error bars are smaller than the symbols.
In all cases where the
comparison was possible we find perfect agreement between the two
methods.


\subsection{9-state model}

In order to demonstrate that our analytic approach leads to the same results 
as other methods of calculation, 
we first reproduce the known results of ref.~\citep{Shuai2007} on opening times and
probabilities in the 9-state model. 
\begin{figure}[htb!]
  \centering
  \subfigure[Total opening probability (\refeq{eq:popenclosed})]{
    \includegraphics[width=.48\textwidth]{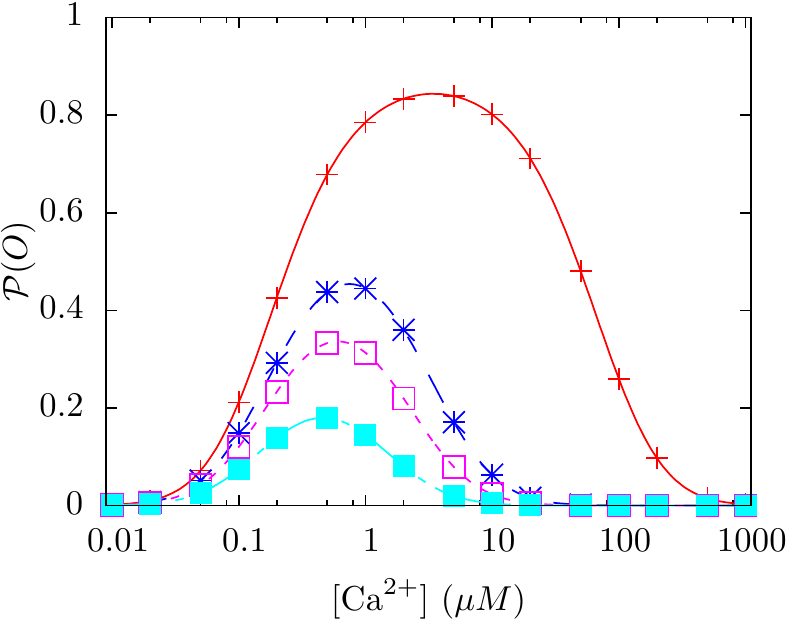}
    \label{fig:popen9s3of4}
  }
  \subfigure[Mean opening time (\refeq{eq:meanopen})]{
    \includegraphics[width=.48\textwidth]{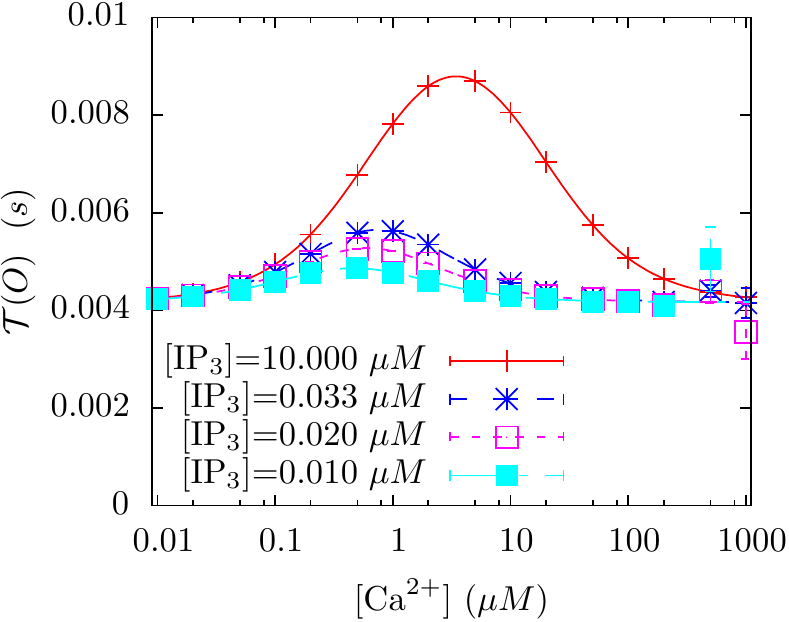}
    \label{fig:tauo9s3of4}
  }
  \caption{Total opening probability and mean opening time for the 9-state model
    for various \ipt concentrations as a function of \ca concentration. The symbols show results from stochastic simulations
    and the solid lines the results obtained from the analytic expressions derived in the text.
    Error bars represent the standard error of the mean along the simulation history.
    Only errors on rare events, which occur at very high \ca concentrations, are visibly
    large in the figures.
  }
\end{figure}
\Reffigures{fig:popen9s3of4} and~\reffign{fig:tauo9s3of4} show the
opening probability and the mean opening
time (\refeqs{eq:popenclosed} and~\refeqn{eq:meanopen}),
respectively. A different analytic derivation
for these observables was provided in ref.~\cite{Shuai2007}.

\begin{figure}[htb!]
  \centering
  \subfigure[Time spent in $L$-related opening (\refeq{eq:tal})]{
    \includegraphics[width=.48\textwidth]{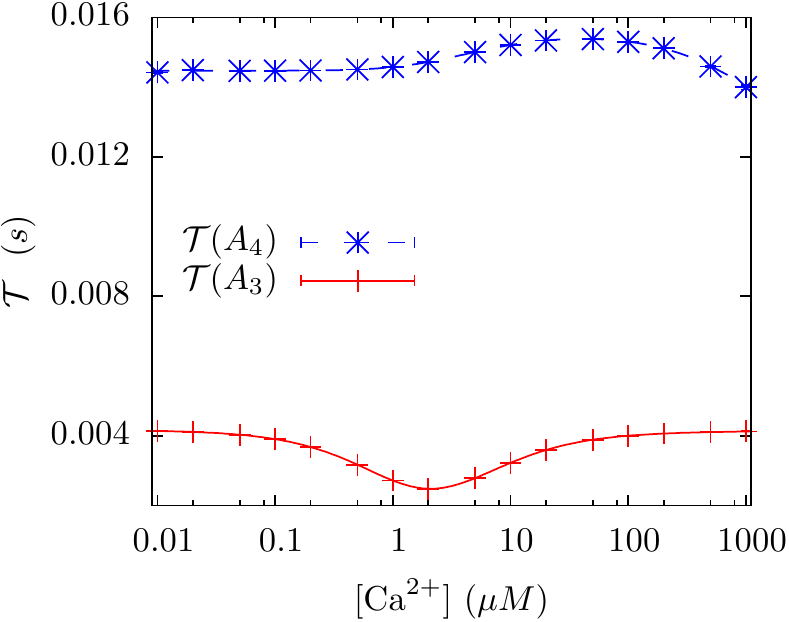}
    \label{fig:TTTF}
  }
  \subfigure[Relative number of opening events (\refeq{eq:pal})]{
    \includegraphics[width=.48\textwidth]{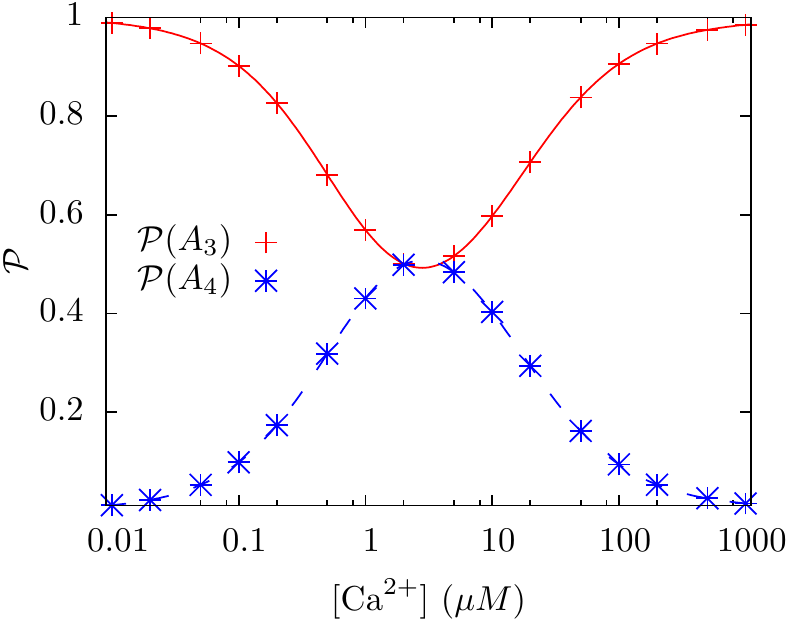}
    \label{fig:PTPF}
  }
  \caption{Statistics for 3- and 4-related opening events for a concentration of ${[\iptm]=10~\mu{}M}$.
    Shown are (a) the mean time spent in 3- and 4-related opening and (b) the relative
    frequencies of these events, determined by stochastic simulations (crosses) and
    through the analytic formulae (solid lines), as a function of \ca concentration.
  }
\end{figure}
\Reffigures{fig:TTTF} and~\reffign{fig:PTPF} show observables associated
to $L$-related opening events. Restricting to the case $N=4$ and $K_{th}=3$, 
we show in particular the time spent in a
$3$- or $4$-related opening event ($\tact{L}$, see \refeq{eq:tal} with $L=3,4$), 
and the probability that an opening event is $3$- or~$4$-related (\refeq{eq:pal}).
For these quantities, analytic expressions have, to the best of our knowledge, not been established previously. 
Thus, these examples demonstrate
the general applicability and usefulness of our approach.
The advantage of analytic expressions of general validity is apparent: Expected times and their variances can be 
computed directly and exactly without the need for stochastic simulations. This is in particular
valuable for rare events where computing times for stochastic simulations can dramatically increase.

\subsection{Comparing different models}

A particular strength of theoretical approaches to biology is the principle ability to suggest
experiments which allow discriminating between conflicting models.
We apply our theory to calculate measurable quantities which display a qualitatively different 
dependency on the external \ca and \ipt concentrations and are therefore in principle 
suitable to differentiate between the three discussed \iptr models.
For the 9-state model we use the parameters as published in~\citet{Shuai2007}.
For the 8-state model we simultaneously fit (see Supplementary
Material~S2) the total opening probability $\popen$ and the opening time
$\topen$ to values obtained from the 9-state model for various \ca
concentrations at ${[\iptm]=10\mu{}M}$ and ${[\iptm]=0.33\mu{}M}$ (the
parameters do not strictly correspond to the best obtained fit but
rather represent a trade off between fitness and reasonable magnitude;
for details see Supplementary Material~S2).
The parameters of the global activation model are the same as the
parameters of the 9-state model, with $\globActRate=a_0$ and
$\globDeactRate=b_0$. All parameters are listed in \reftable{tab:pars}.
\begin{table}
\begin{center}
\begin{tabular}{cccc}
\toprule
  & {9-state}   & {8-state}    & {global} \\
\midrule
  $a_1$ & 60.0  & 56.9338  & 60.0     \\
  $b_1$ & 0.216 & 0.200904     & 0.216    \\
  $a_2$ & 0.2   & 0.190167    & 0.2      \\
  $b_2$ & 3.2   & 13.591   & 3.2      \\
  $a_3$ & 5.0   & 5.0       & 5.0      \\
  $b_3$ & 4.0   & 3.35775    & 4.0      \\
  $a_4$ & 0.5   & 0.5       & 0.5      \\
  $b_4$ & 0.036 & 0.18777    & 0.036    \\
  $a_5$ & 150.0 & 476.698    & 150.0    \\
  $b_5$ & 120.0 & 88.3325    & 120.0    \\
  $a_0$ & 540.0 & & \\
  $b_0$ & 80.0  & & \\
  $\globActRate$  & &       & 540.0     \\
  $\globDeactRate$ & &      & 80.0      \\
\bottomrule
\end{tabular}

\end{center}
\caption{Model parameters used for Gillespie simulation and analytic calculations.}
\label{tab:pars}
\end{table}

Experimentally accessible quantities are the opening and closing times of channels as well as
the inter-spike intervals. 
\begin{figure}[htb!]
  \centering
  \subfigure[Mean opening time (\refeq{eq:meanopen})]{
    \includegraphics[width=.48\textwidth]{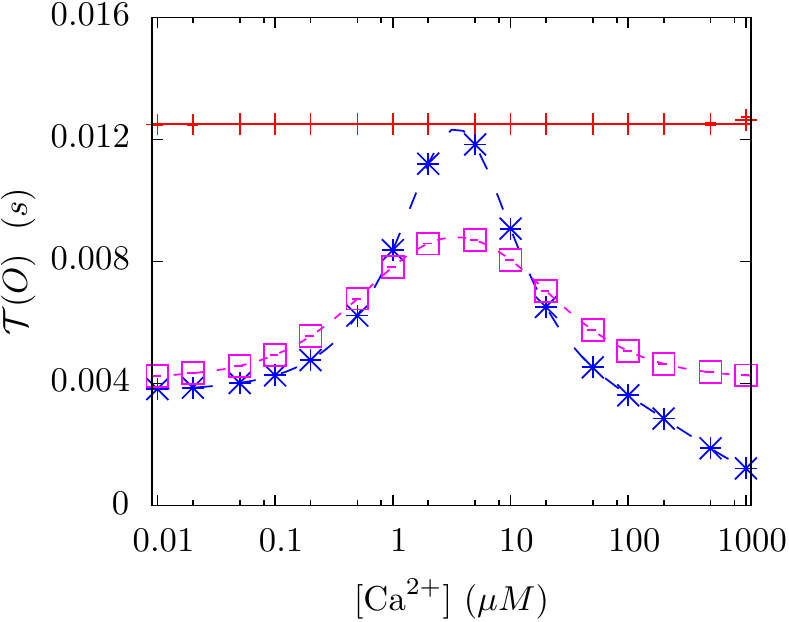}
    \label{fig:tauoAll3of4}
  }
  \subfigure[Mean closing time (\refeq{eq:meanclosed})]{
    \includegraphics[width=.48\textwidth]{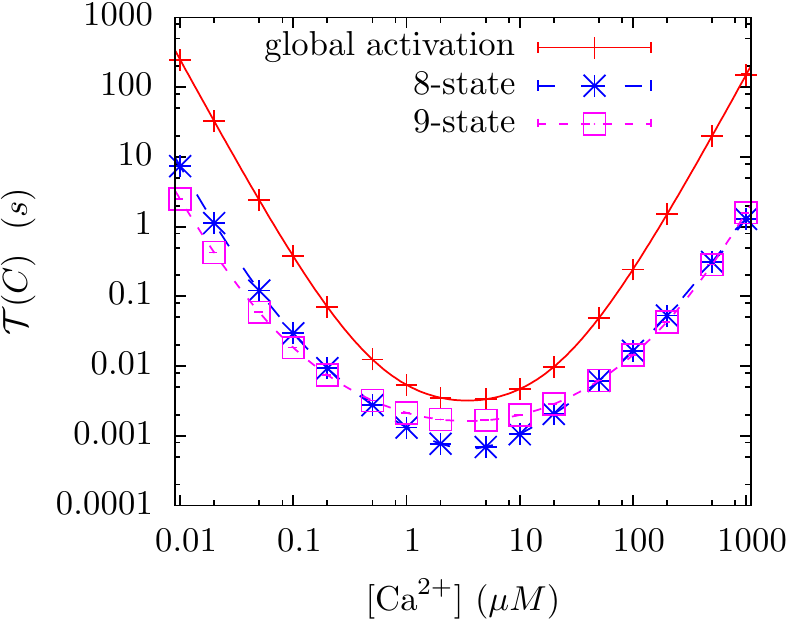}
    \label{fig:taucAll3of4}
  }
  \caption{Comparison of mean opening and mean closing times in the 8-state, 9-state and global activation model for $[\iptm]=10
    \mu{}M$.}
  \label{fig:tauopenclosed}
\end{figure}
In \reffig{fig:tauopenclosed}, the opening and closing times predicted by the three discussed models 
are depicted as a function of the \ca concentration for a fixed value of $[\iptm]$. Whereas the closing times
show qualitatively similar behavior, the opening time predicted by the global activation model
is clearly distinguishable from the 8- and 9-state models.
In
particular, $\topen$ is constant as a function of the \ca concentration for the global activation model,
which can be understood
because the deactivation rate of the open state, $b_{\mathcal{O}}$, is a concentration independent constant. 

Thus, we can conclude that the assumption of global channel activation is
invalid since the predicted constant open time $\topen$ contradicts experimental
measurements which display a bell-shaped dependency on the cytosolic \ca 
concentration~\citep{Bezprozvanny91}. This calculation nicely illustrates the power of our 
approach since it allows to generate functional relations between measurable quantities
and possible hidden mechanisms that can be studied in a parameter dependent manner. 

While opening and closing times alone do not allow to test whether the 8- or the 9-state
model describes experimental data better, this is possible by investigating the
coefficient of variation (CV) of the inter-spike intervals between opening events.
\begin{figure}[htb!]
  \centering
  \subfigure[8-state model]{
    \includegraphics[width=.48\textwidth]{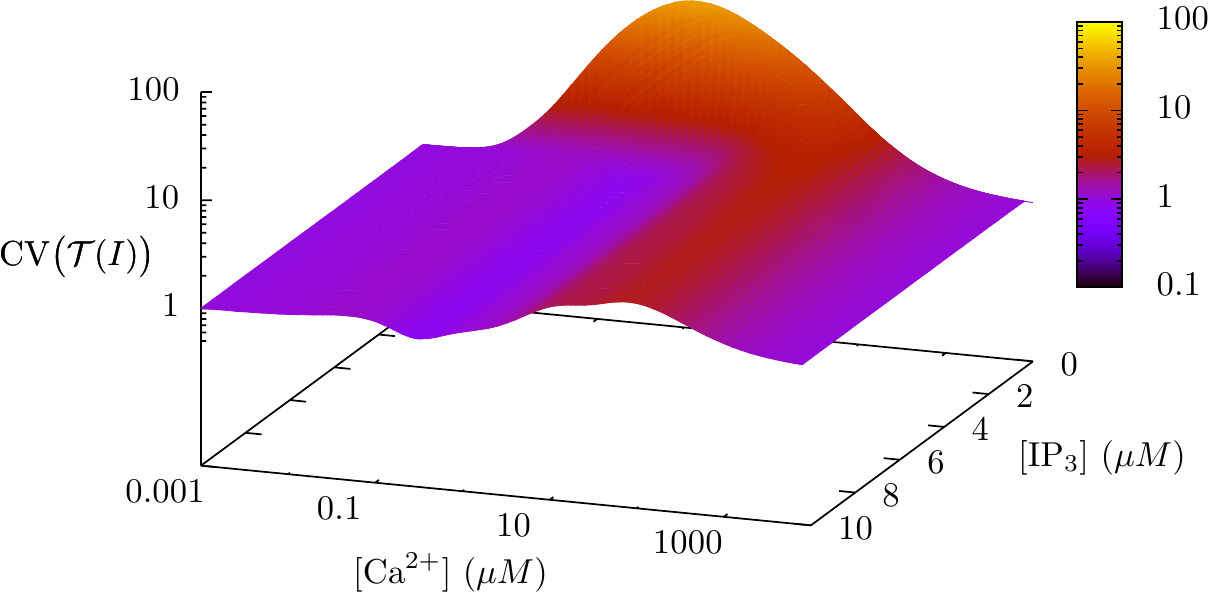}
    \label{fig:isicv8s3of4}
  }\hfill
  \subfigure[9-state model]{
    \includegraphics[width=.48\textwidth]{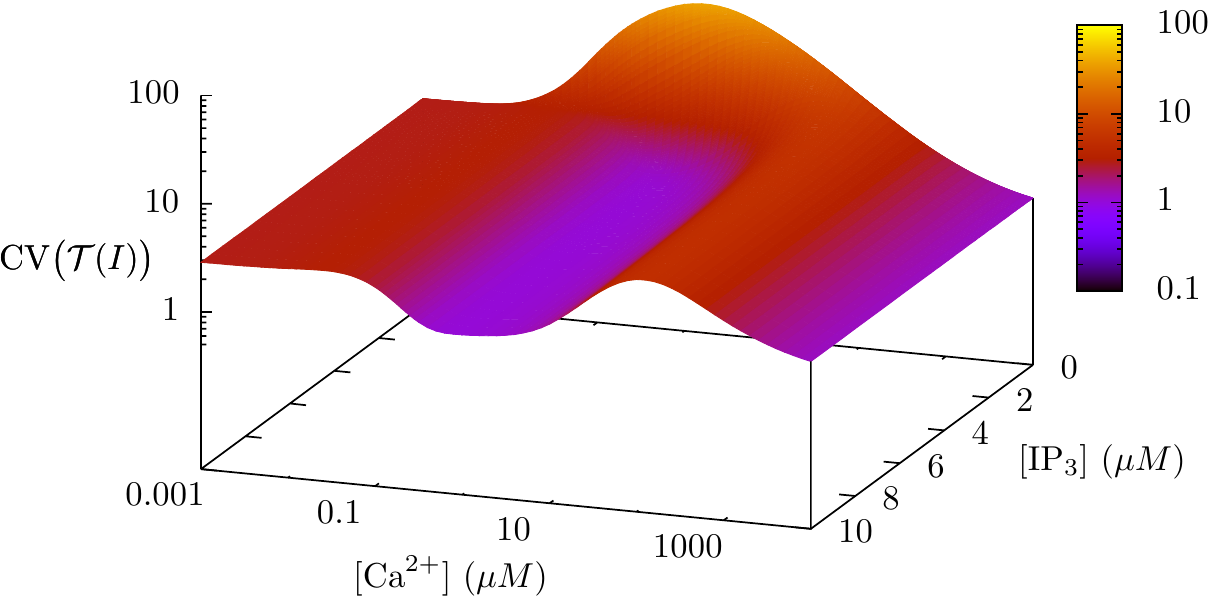}
    \label{fig:isicv9s3of4}
  }
  \caption{Coefficient of variation of inter-spike intervals (\refeq{eq:isicv}) as
    function of \ipt and \ca concentration.}
  \label{fig:ISIcv}
\end{figure}
\Reffigure{fig:ISIcv} shows the functional forms of the CV of the inter-spike intervals
as functions of the \ca and \ipt concentrations.
All models show a qualitatively similar shape, with large changes of
the CV being observed for \ca concentrations between ${[\cam]=0.1\mu{}M}$
and ${[\cam]=10\mu{}M}$, and the largest changes for low \ipt
concentrations with a local maximum around ${[\cam]=2\mu{}M}$.
Only for the case of the 8-state model the base-level of the CV at low
concentrations reaches a value around 1, meaning that the inter-spike
interval distributions collapse to an exponential-like functional form.

This feature may be used to design experiments specifically targeting
low \ca concentrations to distinguish among different models.

\subsection{Receptor channels with an arbitrary number of subunits}

For the \iptr it is established that the internal structure consists
of four subunits~\citep{Mikoshiba2006a, Mikoshiba93,
  Mikoshiba94,Jiang2002}. However, in nature complexes with an almost
arbitrary number of subunits exist~\citep{Unwin1989}.
With the theoretical analysis presented here, it is in principle possible to
distinguish between different models describing the internal structure of a receptor channel.
We consider a generalized 9-state model by assuming that
a channel consisting of $N$ subunits is open if at least $K_{th}$ subunits are in an
active state.
We illustrate the effect of a different internal receptor structure on the dynamic channel
properties for the two cases of a 5-subunit channel which opens with at least 3 active subunits
($N=5$, $K_{th}=3$) 
and a 7-subunit channel which also opens if at least 3 of the subunits are in the activated state
($N=7$, $K_{th}=3$).

\begin{figure}[htb!]
  \centering
  \subfigure[$N=5$, $K_{th}=3$]{
    \includegraphics[width=.48\textwidth]{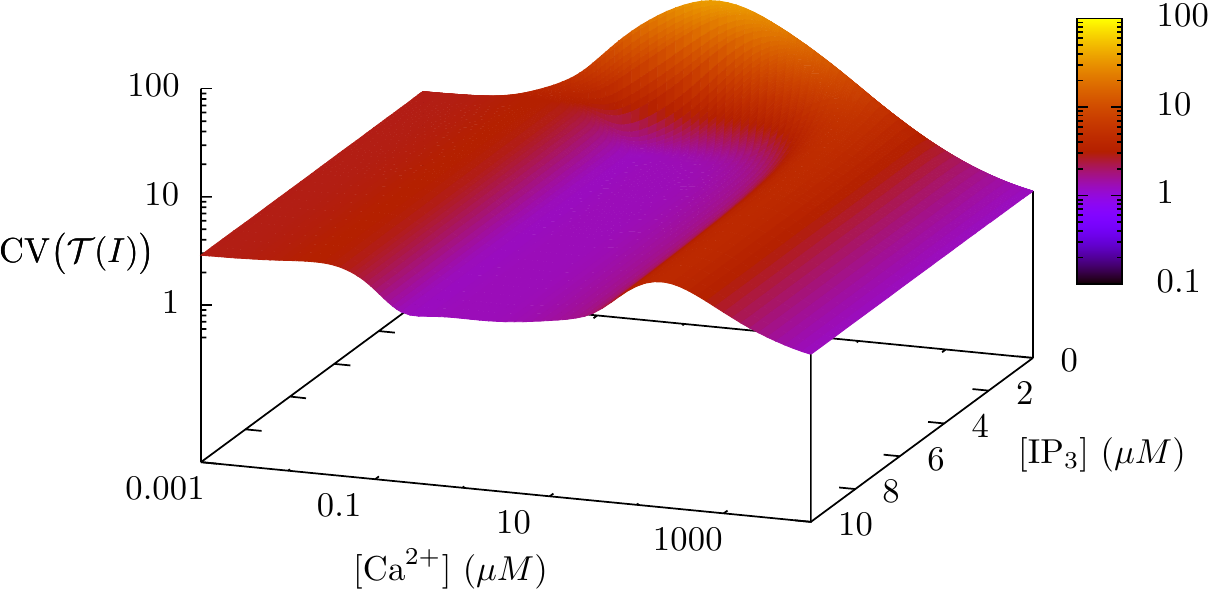}
    \label{fig:isicv8s3of5}
  }
  \subfigure[$N=7$, $K_{th}=3$]{
    \includegraphics[width=.48\textwidth]{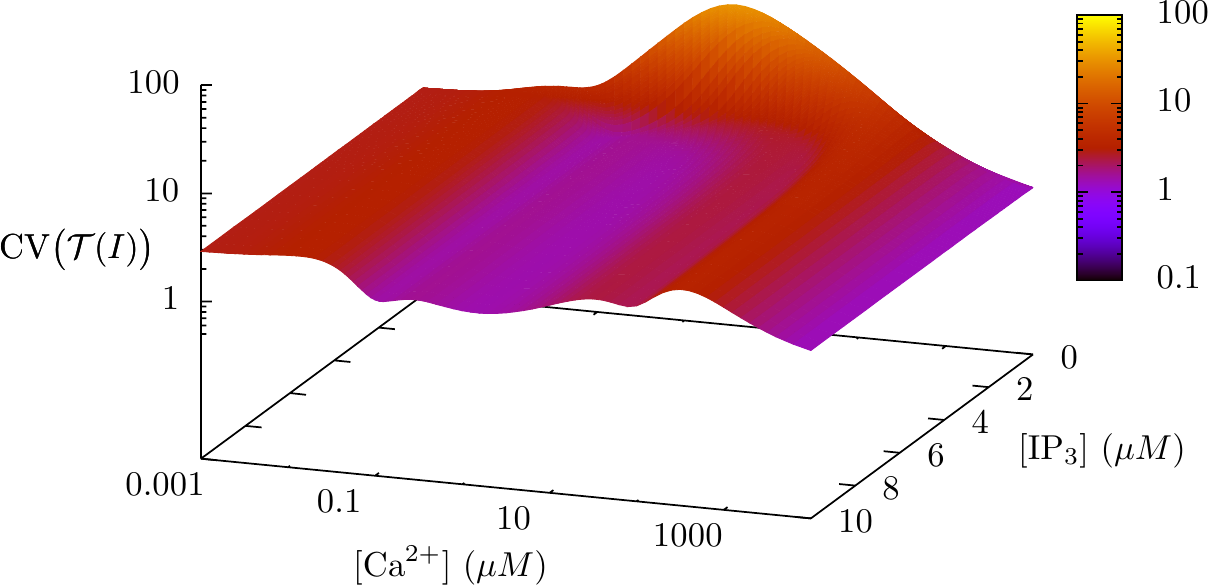}
    \label{fig:isicv9s3of7}
  }
  \caption{Coefficient of variation of inter-spike intervals (\refeq{eq:isicv}) as
    function of \ipt and \ca concentration for models with different
    numbers of subunits.
    Both models assume that a channel opens if at least $K_{th}=3$ subunits are in the
    activated state.
    The results in the left panel are computed for a hypothetical channel with 5 subunits, 
    the right panel for 7 subunits.
  }
  \label{fig:moreSubunitsISIcv}
\end{figure}
The functional form of the CV of the inter-spike intervals are plotted in \reffig{fig:moreSubunitsISIcv}
as functions 
of the \ca and \ipt concentrations for the two hypothetical channel configurations.
The shape of the functions display peculiar features. In particular
in the regime of high \ipt concentrations the CV exhibits an interesting multimodal form for $N=7$. 
This demonstrates the capability of our analytic approach to distinguish between similar models 
which only differ in the internal configuration of the channel.

\section{Conclusions}
\label{sec:conclusions}

The emergence of higher (mesoscopic) levels of organization from a lower level
dynamics is a characteristic feature of complex systems. An
established technique to study mesoscopic properties resulting from
underlying stochastic processes is by detailed numerical
simulations~\citep{Gillespie1977}.
Alternatively, analytical approaches have been developed to determine how
the lower-level processes can lead to observable mesoscopic behaviors such
as bursts of ion channels~\citep{Colquhoun1982}.
These analytical approaches allow in principle for an exact definition
and calculation of transition probability densities and thus for a
fully detailed description of the system properties, as they are based
on the spectral expansion of the transition matrix. Therefore they
rely on solving the corresponding eigenvalue problem and require
following the whole time-evolution of the system.
In this work we have developed an alternative analytic approach allowing for a direct
computation of statistical properties of mesoscopic quantities from the underlying
Markovian dynamics. 
Our methodology provides an attractive computational improvement able to circumvent 
time-consuming stochastic simulations and also to capture rare events in their entirety.
In contrast to other analytical
approaches~\citep{Colquhoun1982}, our method is not able to determine
the full probability density.
However, due to the use of efficient recursive formulae, applicable to
even very large systems, the present approach is able to provide
analytic equations for the computation of the experimental predominant
first statistical moments at the mesoscopic level from elementary
microscopic rate constants.

To illustrate the potential of our general theoretical developments, we
have studied the dynamics of three stochastic models of the so-called \ipt receptor channels,
\ie calcium channels releasing calcium from intracellular stores in a calcium and \ipt dependent
manner. These are of high interest since their local properties shape 
cell-wide \ca signals~\citep{Ruediger2007,Skupin2010,Ruediger2012} and are 
thus responsible for a variety of physiological processes~\citep{Berridge98}. 
In this context, our approach can be seen as a generalized and applicable method to 
predict mesoscopic behavior from underlying complex microscopic dynamics, which was for 
instance explicitly approximated for clustered \ipt receptors~\citep{Higgins2009}.
A slightly different approach for characterizing the dynamics of such
a mesoscopic system was recently developed that is based on waiting
time distributions of observable mesoscopic
variables~\citep{Moenke2012}. Due to the non-Markovian character of the
mesoscopic dynamics, the considered waiting-time probabilities are
non-exponential making calculations of the statistical moments of the
dynamics based on Laplace transformation potentially
problematic. Using reasonable approximations for the channel
waiting-time distributions, the approach, however, enables direct and fast
simulations of \ca dynamics that fits experimental results nicely even
though it does not make reference to a mechanistic molecular channel model.

We demonstrate the general applicability of our theory by deriving analytic
formulae for mesoscopic quantities which previously were only accessible through 
stochastic simulations. These quite complex mesoscopic quantities, corresponding to 
channel opening events in which only a certain number of subunits are activated,
are illustrative for understanding how the average channel opening times and probabilities
are governed by the molecular structure of the receptors.

As a key application of our theoretical concepts we systematically investigate 
alternative models of the \ipt receptor channels and predict experimentally 
accessible mesoscopic quantities which can be used to discriminate between
the models and establish which mechanistic model best describes
the experimental observations.
In particular, we have shown that the assumption of the global activation 
mechanism contradicts experimental data because it predicts a \ca independent 
open time whereas a bell-shaped dependency is observed in experiments.
Such ability to distinguish between competing mechanistic models is a key motivation for
theoretical approaches to biology in general.
A particular strength of analytic formulae is
that they automatically capture rare events, which in stochastic simulations may easily be missed
or are only accurately described with extremely long simulations.
However, precisely these rare events are at the basis of the intermittency properties~\citep{Frisch1996} of
complex dynamic systems and neglecting rare events may result in an inaccurate description of
the dynamic properties of the investigated system.

While the investigated models all represent closed systems without external fluxes
into or out of the system, our approach is applicable also to open systems
in stationary state.
The only assumption that enters our theoretical approach is that of stationarity,
which means that the sum of the probability rates leading to a microstate should equal the
sum of the probability rates of processes leaving that state, see \refeq{eq:SUMR}.
For the calculation of all quantities which do not depend on stationary probabilities,
including all those which are conditional on the residence of the system in a particular
microstate,
no modification is necessary for open systems. 
Those quantities depending on the stationary probability distribution also keep their functional form,
but the presence of external fluxes requires a modification of the calculation of the
stationary probabilities~(\refeq{eq:LINEQ}).
Let $v_n^\mathrm{in}$ and $v_n^\mathrm{out}$ denote the external fluxes into and out of microstate $n$,
respectively. For the fluxes leaving the system, we can assign rate constants analogous to
\refeq{eq:VIJ} and write
\begin{equation}
  v_n^\mathrm{out}=\pei{n}k_n^\mathrm{out}\, , \label{eq:VIout}
\end{equation}
leading to the balance equation
\begin{equation}
 v_n^\mathrm{in} + \sum\limits_{m} p_m k_{mn}  = p_n \left( k_n^\mathrm{out}+\sum\limits_{l} k_{nl} \right) \, , \notag
\end{equation}
from which the stationary, non-equilibrium, probabilities can be determined by solving the
inhomogeneous linear system of equations
\begin{equation}
\sum\limits_{m} p_m \left(k_{mn} -  \delta_{mn}\left( k_m^\mathrm{out}+\sum\limits_\ell k_{m\ell}\right)\right) = 
-v_n^\mathrm{in} \, .\label{eq:LINEQINHOM}
\end{equation}

The derivation of the expressions for second moments of the dwell time distributions
was based on the assumption that the underlying Markovian microscopic process is
a Poisson process. The same assumption is fundamental for the Gillespie algorithm~\citep{Gillespie1977}
on which stochastic simulations are based. For atomic and molecular interactions, this is a reasonable assumption
which also underlies the derivation of the mass-action kinetic rate law for chemical
reactions. Under this condition, there is no principle obstacle to derive expressions for the
higher moments of the dwell time distribution in mesostates. With such expressions available
the distribution
of residence times can be approximated by analytic expressions with in principle arbitrary accuracy.
The limitations of the applicability of this approach will probably lie in an increasing complexity
of the resulting expressions and a concomitant difficulty of their numerical evaluation.
However, even only with the analytic expressions for first and second moments presented in this work,
the possibility has been established to extract relevant information about mesoscopic and experimentally accessible quantities
for model discrimination and refinement.

\section*{Acknowledgments}
\label{sec:acknowledgments}
The authors would like to express their gratitude to Prof.~Reinhart
Heinrich, who was the focal point of our collaboration but sadly
deceased far too early.

They also wish to thank Prof. Michael Pusch for
an enlightening discussion.

NC, KJ, SM and GCR acknowledge the Galileo Galilei Institute for
Theoretical Physics for the hospitality offered to them, in the
occasion of the workshop "New Frontiers in Lattice Gauge Theories",
while this work was brought to an end.
GCR would like to thank MIUR (Italy) for partial support under the
PRIN contract number 20093BMNPR. SM would like to thank MIUR (Italy)
for partial support under the PRIN contract number 20083Y4Y7.
OE and NC were supported by the Scottish Funding Council
through the Scottish Universities Life Science Alliance (SULSA).
NC and AS are supported by the knowledge transfer program by the
health transfer initiative of the Luxembourg government.

\section*{Author contributions}
\label{sec:authorcontr}
All authors designed the study during scientific discussions.
NC implemented the algorithms and performed all numerical simulations.
OE had the original idea of history-dependent mesostate characteristics.
NC, OE and GCR performed the analytic calculations.
AS provided support in code development, contributed to model development and supported
analytic calculations.
AS and SM were instrumental in guiding the project towards biological relevance.
KJ provided expertise in the mathematical and statistical formalisms.
All authors wrote the paper.

\newpage{}
\section*{Supplementary Material}

\newpage{}
\setcounter{section}{0}
\setcounter{subsection}{0}
\setcounter{subsubsection}{0}
\setcounter{equation}{0}
\renewcommand{\thesection}{S1.\arabic{section}}
\renewcommand{\thesubsection}{S1.\arabic{subsection}}
\renewcommand{\thesubsubsection}{S1.\arabic{subsubsection}}
\renewcommand{\theequation}{S1.\arabic{equation}}

\section{Detailed derivation of $\QUV{U}{V}{i}{k}$}
\label{suppsec:quvik}

In this section we derive in detail the formula of the matrix elements
$M_{ij}$ (\refeqmain{eq:piiinUjinUcondUV}), which describes the
process of visiting another state $j\ne i$ within $U$ before leaving
to $V$.

We first calculate the
probability that the system leaves directly to any microstate of $V$,
under the condition that from $i \in U$ the next mesostate
is $V$, by summing over $k$ in \refeqmain{eq:piiinUkinVcondUV}
\begin{equation}
  \label{suppeq:P3direct}
  \frac{\sum\limits_{k' \in V}\piij{ik'}}{\PiUV{U}{V}{i}} \, .
\end{equation}
So, conversely, the conditional probability that the system first visits
any other microstate in $U$, and later leaves to mesostate $V$, is
\begin{equation}
  \label{suppeq:pnextinUgivenUtoV}
  1 - \frac{\sum\limits_{k' \in V}\piij{ik'}}{\PiUV{U}{V}{i}}\, .
\end{equation}
Given that the next mesostate transition is $U \rightarrow V$, but
first a microstate in $U$ is visited, the conditional probability
that this microstate is $j \in U$ is
\begin{equation}
  \label{suppeq:pnextinUisjGivenUtoVandNextIsU}
  \frac{\piijstarUUV{i}{j}{U}{V}}{\sum\limits_{i' \in U} \piijstarUUV{i}{i'}{U}{V}}
  = \frac{\piij{ij}\PiUV{U}{V}{j}}{\sum\limits_{i' \in U} \piij{ii'}\PiUV{U}{V}{i'}}\, .
\end{equation}
Altogether this leads to
\begin{equation}
  \label{suppeq:piiinUjinUcondUV}
  M_{ij}= \piiinUjinUcondUV{i}{j}{U}{V}= \left(1 - \frac{\sum\limits_{k' \in V}\piij{ik'}}{\PiUV{U}{V}{i}}\right)
  \frac{\piij{ij}\PiUV{U}{V}{j}}{\sum\limits_{i' \in U} \piij{ii'}\PiUV{U}{V}{i'}} \, .
\end{equation}

\section{Derivation of more complex dwell times}
\paragraph{Time spent in mesostate $U$ when the system is in $i \in U$ and in the next mesostate transition it arrives in $k \in V$}
This quantity differs from \refeqmain{eq:tUcondiinUandUtoV} by the
requirement that the system arrives in a specific microstate $k \in
V$, therefore the matrix elements needed for its derivation are
different. Apart from that the derivation is similar and one obtains
\begin{align}
  & \WUV{U}{V}{i}{k} = \taui{i} \notag \\
  & \ + \sum\limits_{j \in U}
  \underbrace{\piiinUjinUcondUkinV{i}{j}{U}{k}{V}}_{N_{ij}} \WUV{U}{V}{j}{k}  \notag\\
  & \WUV{U}{V}{i}{k} = \sum\limits_{j \in U}\left(\mathbf{1} - \mathbf{N}\right)^{-1}_{ij} \taui{j} \, ,
\end{align}
where the matrix elements $N_{ij}$ are
\begin{align}
  \piiinUjinUcondUkinV{i}{j}{U}{k}{V}=
  \left[\left(1 - \frac{\piij{ik}}{\RUV{U}{V}{i}{k}}\right)
      \frac{\piij{ij}\RUV{U}{V}{j}{k}}{\sum\limits_{i' \in U} \piij{ii'}\RUV{U}{V}{i'}{k}} \right]
\end{align}
For the corresponding second raw moment one then gets under the assumption of exponentially distributed $\taui{i}$
\begin{align}
  \WUVsq{U}{V}{i}{k} &= 2 \sum\limits_{j \in U}\left(\mathbf{1} - \mathbf{N}\right)^{-1}_{ij} \taui{j} \, \WUV{U}{V}{j}{k} \, .
\end{align}

\paragraph{Time spent in mesostate $U$ given that the mesostate
  transition chain is $W \rightarrow U \rightarrow V$}
The formulae derived in the main text allow us to calculate
probabilities and dwell times of arbitrary complicated transition
chains. For example, the time spent in mesostate $U$, given that the
mesostate transition chain is $W \rightarrow U \rightarrow V$, is
\begin{equation}
  \label{suppeq:zwuv}
  \ZWUV{W}{U}{V}=\sum\limits_{j \in U} \CWUV{W}{U}{V}{j} \  \XUV{U}{V}{j} \, ,
\end{equation}
where
\begin{equation}
  \label{suppeq:cwuv}
  \CWUV{W}{U}{V}{j}=\frac{\AWU{W}{U}{j} \, \PiUV{U}{V}{j}}{\sum\limits_{j' \in U} \AWU{W}{U}{j'} \, \PiUV{U}{V}{j'}} 
\end{equation}
is the probability that in the chain $W \rightarrow U \rightarrow V$
the first visited microstate in $U$ is $j$.

The corresponding second raw moment is calculated by weighting each
$\XUVsq{U}{V}{i}$ with its probability (\refeq{suppeq:cwuv})
\begin{equation}
  \ZWUVsq{W}{U}{V}=\sum\limits_{j \in U} \CWUV{W}{U}{V}{j} \  \XUVsq{U}{V}{j} \, .
\end{equation}

Similar to \refeqmain{eq:P1}, $\ZWUV{W}{U}{V}$ and $\ZWUVsq{W}{U}{V}$ are
conditional only on mesostates (not on a particular microstate) and
therefore depend on the stationary probabilities.

\paragraph{Time spent in mesostate $U$ when system is in $i \in U$ given that the mesostate
  transition chain is $U \rightarrow V \rightarrow Z$}
For a transition chain with more than one mesostate visited ``in the
future'', the matrix that needs to be inverted changes, but the
derivation follows the same procedure as used when deriving simpler
dwell times. We will derive the dwell time in $U$ where the next
mesostates will be $V$ and $Z$
\begin{equation}
  \XUVY{U}{V}{Z}{i} = \taui{i}  + \sum_{j \in U} \piiinUjinUcondUVZ{i}{j}{U}{V}{Z} \ \XUVY{U}{V}{Z}{j} \, ,
\end{equation}
leading to
\begin{align}
  \label{suppeq:xuvy}
  & \XUVY{U}{V}{Z}{i} = \sum_{j \in U}\left(\mathbf{1} - \mathbf{M}^{UVZ}\right)^{-1}_{ij} \taui{j} \, ,
\end{align}
where the matrix elements are $M^{UVZ}_{ij}=\piiinUjinUcondUVZ{i}{j}{U}{V}{Z}$.
Similar to \refeqmain{eq:piiinUjinUcondUV}, these can be calculated in the following way
\begin{align}
  \label{suppeq:piiinUjinUcondUVZ}
  & \piiinUjinUcondUVZ{i}{j}{U}{V}{Z} \notag \\
  & \qquad = \left(1 - \frac{\sum_{k' \in V} \piijstarUVZ{i}{k'}{U}{V}{Z}}{\PiUVZ{U}{V}{Z}{i}} \right)
  \frac{\piijstarUUVZ{i}{j}{U}{V}{Z}}{\sum_{j' \in U} \piijstarUUVZ{i}{j'}{U}{V}{Z}}  \notag \\
  & \qquad =\left(1 - \frac{\sum_{k' \in V} \piij{ik'} \PiUV{V}{Z}{k'}}{\PiUVZ{U}{V}{Z}{i}} \right)
  \frac{\piij{ij} \PiUVZ{U}{V}{Z}{j}}{\sum_{j' \in U} \piij{ij'} \PiUVZ{U}{V}{Z}{j'}} \, ,
\end{align}
where the probability that from $i \in U$ the mesostates $V$ and $Z$ are visited is
\begin{align}
  \PiUVZ{U}{V}{Z}{i} &= \sum_{k' \in V} \RUV{U}{V}{i}{k'} \PiUV{V}{Z}{k'} \, .
\end{align}

\paragraph{Time spent in mesostate $U$ given that the mesostate
  transition chain is $W \rightarrow U \rightarrow V \rightarrow Z$}
Here we will derive the dwell time in $U$ when the
previous mesostate was $W$ and the next will be $V$ and then
$Z$. Similar to \refeq{suppeq:zwuv}, we write this as
\begin{align}
  \ZWUVYinU{W}{U}{V}{Z}=\sum_{j \in U} \PWUVYiinU{W}{U}{V}{Z}{j} \ \XUVY{U}{V}{Z}{j} \, ,
\end{align}
where the first term corresponds to \refeq{suppeq:cwuv}
\begin{align}
  \PWUVYiinU{W}{U}{V}{Z}{j} &= \frac{\AWU{W}{U}{j} \PiUVZ{U}{V}{Z}{j}}{\sum_{j' \in U} \AWU{W}{U}{j'} \PiUVZ{U}{V}{Z}{j'}} \, ,
\end{align}
and the second term is the dwell time derived in the previous paragraph (\refeq{suppeq:xuvy}).

The examples presented above illustrate how an arbitrary complex
quantity can be expressed by quantities of lower complexity.

\section{Derivation of more complex quantities for mesostate subsets}

For the calculations in \refsecmain{sec:iptrmodels},
we require
the time spent in mesostate $U$ given that the mesostate transition
chain is ${W {\rightarrow} U {\rightarrow} \firstinbarVfromV{V'}{V}
  {\rightarrow} Z}$, which is given by
\begin{align}
  \label{suppeq:generictalPhase1}
  & \ZWUVbarVYinU{W}{U}{V'}{V}{Z} = \sum\limits_{j \in U}
   \PWUVbarVYiinU{W}{U}{V'}{V}{Z}{j} \XUVY{U}{\firstinbarVfromV{V'}{V}}{Z}{j} \, .
\end{align}
The probability which enters \refeq{suppeq:generictalPhase1} takes the form
\begin{align}
  \label{suppeq:genericprobPhase1}
  & \PWUVbarVYiinU{W}{U}{V'}{V}{Z}{j}=
  \dfrac{\AWU{W}{U}{j} \ \PsiUVbarVY{U}{V'}{V}{Z}{j}}
  {\sum\limits_{j' \in U} \AWU{W}{U}{j'} \ \PsiUVbarVY{U}{V'}{V}{Z}{j'}} \, .
\end{align}
The matrix elements
$M^{U\firstinbarVfromV{V'}{V}Z}_{ij}=\piiinUjinUcondUVZ{i}{j}{U}{\firstinbarVfromV{V'}{V}}{Z}$
needed for the derivation of
$\XUVY{U}{\firstinbarVfromV{V'}{V}}{Z}{i}$ can be calculated in analogy to \refeq{suppeq:piiinUjinUcondUVZ}.
For the calculation, the transition chains $U {\rightarrow} V {\rightarrow} Z$ in the probabilities have to be replaced
by the chains $U {\rightarrow} \firstinbarVfromV{V'}{V} {\rightarrow} Z$. These probabilities are defined in \refeqmain{eq:Pmesostatesubset}.

\newpage{}
\setcounter{section}{0}
\setcounter{subsection}{0}
\setcounter{subsubsection}{0}
\setcounter{equation}{0}
\renewcommand{\thesection}{S2.\arabic{section}}
\renewcommand{\thesubsection}{S2.\arabic{subsection}}
\renewcommand{\thesubsubsection}{S2.\arabic{subsubsection}}
\renewcommand{\theequation}{S2.\arabic{equation}}

\section{Fitting 8-state model parameters}
\label{supp2sec:fitting}
\newcommand{\weightpopen}{W_{\popen}}
\newcommand{\weighttopen}{W_{\topen}}
\newcommand{\Plimits}{\mathcal{L}_{\textrm{imits}}(p^{\textrm{8-state}})}
To determine parameters for the 8-state model
($p^{\textrm{8-state}}$), we have first calculated the total opening probability
$\popen$ and the opening time $\topen$ of the 9-state model at various
\ca and \ipt concentrations using the parameters
$p^{\textrm{9-state}}$ from~\citet{Shuai2007}, and used these values as
input for a fitting procedure.

As the two observables $\popen$ and $\topen$ do not depend on all
parameters, the fitting procedure was reduced to the parameters $a_1,
b_1, a_2, b_2, b_3, b_4, a_5, b_5$, with the parameters $a_3=5$ and
$a_4=0.5$ held fixed.

To have equally logarithmically spaced calcium concentrations we took
$[\cam]={10^{x}\mu{}M}$ with
${x=-0.7, -0.6, \dots, 1.9, 2.0}$, leading to concentrations in the
range ${[\cam]=[0.2,100] \mu{}M}$. For the \ipt concentrations we
have chosen ${[\iptm]=10\mu{}M}$ and ${[\iptm]=0.33\mu{}M}$.

We used the genetic optimization package rgenoud~\citep{Mebane2011} to
minimize a $\chi^2$~function where the $\popen$ and $\topen$
contributions are weighed with the coefficients $\weightpopen$ and
$\weighttopen$
\begin{align}
  \chi^2=& \weightpopen \sum_{\cam, \iptm}\Big(\popen\big(p^{\textrm{9-state}},\cam,\iptm\big) - \popen\big(p^{\textrm{8-state}},\cam,\iptm\big) \Big)^2 \notag \\
  &+ \weighttopen \frac{1}{s} \sum_{\cam, \iptm}\Big(\topen\big(p^{\textrm{9-state}},\cam,\iptm\big) - \topen\big(p^{\textrm{8-state}},\cam,\iptm\big) \Big)^2 \notag
  \ .
\end{align}
Note that $\frac{1}{s}$ is required to make the contribution of the
opening times dimensionless.

We arbitrarily chose $\weightpopen = \weighttopen = 0.5$.
Depending on the given random seed, the fitted parameters differ
dramatically while having compatible values of $\chi^2$, indicating
that the fitness landscape is rather flat but \emph{bumpy}.
We therefore decided to restrain all parameters between $0.01$
and~$500$ (using genoud's \emph{Domains} parameter) to look for
parameters of reasonable size, and randomly chose one of the
solutions.

We should emphasize here that this result does not correspond to a
proper fit, but rather represents a solution of $\chi^2$-minimization
leading to reasonable values of the parameters.

\newpage{}
\bibliographystyle{model2-names}
\bibliography{references}

\end{document}